\documentclass[usenatbib]{mn2e}
\usepackage{graphicx}
\newcommand{\erg}{\,\mbox{erg}}
\newcommand{\cm}{\,\mbox{cm}}
\newcommand{\s}{\,\mbox{s}}
\newcommand{\yr}{\,\mbox{yr}}
\newcommand{\sr}{\,\mbox{sr}}
\newcommand{\Gyr}{\,\mbox{Gyr}}
\newcommand{\Mpc}{\,\mbox{Mpc}}
\newcommand{\kpc}{\,\mbox{kpc}}
\newcommand{\kms}{\,\mbox{km}\,\mbox{s}^{-1}}
\newcommand{\Lsun}{\,L_{\sun}}
\newcommand{\Msun}{\,M_{\sun}}
\newcommand{\msun}{\,M_{\sun}}

\newcommand{\bgunit}{\erg \s^{-1} \cm^{-2} \sr^{-1}}
\newcommand{\rhostar}{\rho_\star}
\newcommand{\rhorem}{\rho_{\star {\it rem}}}
\newcommand{\rhoform}{\rho_{\star {\it form}}}
\newcommand{\rhodotform}{\dot{\rho}_{\star {\it form}}}

\newcommand{\sfr}{{\it SFR}}
\newcommand{\tsim}{\sim\!}
\newcommand{\ea}{et al.~}
\newcommand{\NN}{\mathcal{N}}

\newcommand{\jbg}{J_{\it EBL}}
\newcommand{\gtrsim}{\ga}
\newcommand{\lesssim}{\la}
\newcommand{\aap}{A\&A}
\newcommand{\araa}{ARA\&A}
\newcommand{\apj}{ApJ}
\newcommand{\apjl}{ApJ}
\newcommand{\apjs}{ApJS}
\newcommand{\mnras}{MNRAS}
\newcommand{\aj}{AJ}
\newcommand{\apss}{Ap\&SS}
\newcommand{\pasp}{PASP}
\newcommand{\pasj}{PASJ}
\newcommand{\nat}{Nat}
\newcommand{\nodata}{$\cdots$}  

\title{On the Evolutionary History of Stars and their Fossil Mass and Light}
\author[Mark A. Fardal et al.]{Mark A. Fardal$^1$, Neal Katz$^1$,    
    David H. Weinberg$^2$, and Romeel Dav\'e$^3$ \\
  $^1$Department of Astronomy, University of Massachusetts, 
      Amherst, MA, 01003, USA \\
  $^2$Department of Astronomy, Ohio State University, Columbus, OH 43210, USA \\
  $^3$Steward Observatory, University of Arizona, Tucson, AZ, 85721, USA}
\date{draft version \today}
\begin{document}
\maketitle  
\begin{abstract}
  The total extragalactic background radiation can be an important
  test of the global star formation history (SFH).  Using direct
  observational estimates of the SFH, along with standard assumptions
  about the initial mass function (IMF), we calculate the total
  extragalactic background radiation and the observed stellar density
  today.  We show that plausible SFHs allow a significant range in
  each quantity, but that their {\it ratio} is very tightly
  constrained.  Current estimates of the stellar mass and
  extragalactic background are difficult to reconcile, as long as the
  IMF is fixed to the Salpeter slope above $1 \msun$.  The joint
  confidence interval of these two quantities only agrees with that
  determined from the allowed range of SFH fits at the $3\sigma$
  level, and for our best-fit values the discrepancy is about a factor
  of two.  Alternative energy sources that contribute to the
  background, such as active galactic nuclei (AGN), Population III
  stars, or decaying particles, appear unlikely to resolve the
  discrepancy.  However, changes to the IMF allow plausible solutions
  to the background problem.  The simplest is an average IMF with an
  increased contribution from stars around 1.5--$4 \msun$.  A
  ``paunchy'' IMF of this sort could emerge as a global average if low
  mass star formation is suppressed in galaxies experiencing rapid
  starbursts.  Such an IMF is consistent with observations of
  star-forming regions, and would help to reconcile the fossil record
  of star formation with the directly observed SFH.
\end{abstract}

\newcommand{\imftable}{
\begin{table}
\begin{minipage}{72mm}
\caption{Initial Mass Functions}
\label{table.imf} 
\begin{tabular}{lcccc}
\hline
Form$^a$ & $K$ & $x$ & $M_l$ & $M_u$ \\
\hline
Salpeter     & 0.172 & 1.35 & 0.1 & 100 \\
diet Salpeter& 0.218 & 1.35 & 0.188 & 100 \\
Kennicutt    & 0.328 & 0.4  & 0.1 & 1 \\
             & 0.328 & 1.5  & 1   & 100  \\
Miller-Scalo & 0.354 & 0.4  & 0.1 & 1 \\
             & 0.354 & 1.5  & 1   & 10  \\
             & 2.02  & 2.3  & 10  & 100  \\
Kroupa 1993  & 0.579 & 0.3  & 0.1 & 0.5  \\
             & 0.310 & 1.2  & 0.5 & 1  \\
             & 0.310 & 1.7  & 1   & 100  \\
Chabrier     & \nodata & \nodata  & 0.1 & 1  \\
             & 0.238 & 1.3  & 1 & 100  \\
Kroupa 2001  & 0.449 & 0.3  & 0.1 & 0.5  \\
             & 0.224 & 1.3  & 0.5 & 100  \\
Baldry \& Glazebrook  & 0.323 & 0.5  & 0.1 & 0.5  \\
             & 0.199 & 1.2  & 0.5 & 100  \\
Paunchy      & 0.315 & 0    & 0.1 & 0.5 \\
             & 0.194 & 0.7  & 0.5 & 4.0  \\
             & 0.676 & 1.6  & 4.0 & 100  \\
Obese        & 0.083 & 1.35 & 0.188 & 100 \\
             & 0.905 & 1.65 & 1.5 & 100  \\
Extreme top-heavy & 0.136 & 0.95  & 0.1 & 100 \\
\hline
\end{tabular}
$^a$ The IMF is written as ${d\NN}/{d\ln M} = K ({M}/{\msun})^{-x}$,
for $M_l < M < M_u$.  The Chabrier IMF has a lognormal form 
$dN/d\ln(M) \propto \exp[-(\log(M/0.08))^2/(2 \cdot 0.69^2)]$ below
$1 \msun$, and is continuous at $1 \msun$.  
In some cases the original versions of the IMF continued to lower mass, 
but we have imposed a uniform lower mass limit of $0.1 \msun$.
\end{minipage}
\end{table}
}


\newcommand{\localmasstable}{
\begin{table*}
\centering
\begin{minipage}{100mm}
\caption{Local Stellar Mass and Light (for diet Salpeter IMF)}
\label{table.localmass} 
\begin{tabular}{lll}
\hline
\hline
$\rho_z^{a}$  & $1.93 \pm 0.20$  & \citet{bell03} \\
\hline
$\rho_J^{b}$  & $2.11 \pm 0.31$  & \citet{cole01} \\
             & $2.43 \pm 0.08$  & \citet{eke05} \\
             & $2.22 \pm 0.15$  & \citet{jones06} \\
weighted avg & $2.37 \pm 0.36$  & (assuming 15\% systematic error) \\
\hline
$\rho_K^{c}$  & $4.24 \pm 0.64$  & \citet{cole01} \\
             & $4.08 \pm 0.09$  & \citet{bell03} \\
             & $4.66 \pm 0.15$  & \citet{eke05} \\
             & $4.30 \pm 0.37$  & \citet{jones06} \\
weighted avg & $4.24 \pm 0.64$  & (assuming 15\% systematic error) \\
\hline
$\rhorem$    & $3.94 \pm 0.58$  & \citet{cole01} \\
             & $3.71 \pm 0.07$  & \citet{bell03} \\
             & $3.26 \pm 0.11$  & \citet{panter04} \\
             & $3.08 \pm 0.14$  & \citet{eke05} \\
unweighted avg & $3.50 \pm 0.70$  & (assuming 25\% systematic error) \\
\hline
$^{a}$Using $M_{\sun(z)}=4.53$ \citep{bell03}\\
$^{b}$Using $M_{\sun(J)}=3.33$ \citep{worthey94}\\
$^{c}$Using $M_{\sun(K)}=3.70$ \citep{worthey94}\\
\end{tabular}
\end{minipage}
\end{table*}
}


\newcommand{\sfrconvtable}{
\begin{table*}
\centering
\begin{minipage}{100mm}
\caption{Conversion Factors for Star Formation Rate Indicators}
\label{table.sfrconv} 
\begin{tabular}{lrrrcccc}
\hline
IMF     & 
$10^7$     &
$10^8$ & 
$10^9 \yr$ & 
$10^7$ &
$10^8$ & 
$10^9 \yr$ &
$(10^{41} \mbox{erg} \s^{-1} \Msun^{-1} \yr)$ \\
\hline
Salpeter  & 4.31 &  6.51 & 8.60 & 5.00 & 7.68 & 8.76 & 1.22 \\ 
Diet Salpeter & 5.52 &  8.33 & 11.0 & 6.35 & 9.75 & 11.1 & 1.55 \\
Kennicutt &  4.94 &  8.05 & 11.50 & 5.78 & 9.64 & 11.4 & 1.30 \\
\hline
\end{tabular}
\end{minipage}
\end{table*}
}

\newcommand{\obsbackgdfig}{
\begin{figure*}
\includegraphics[width=14cm]{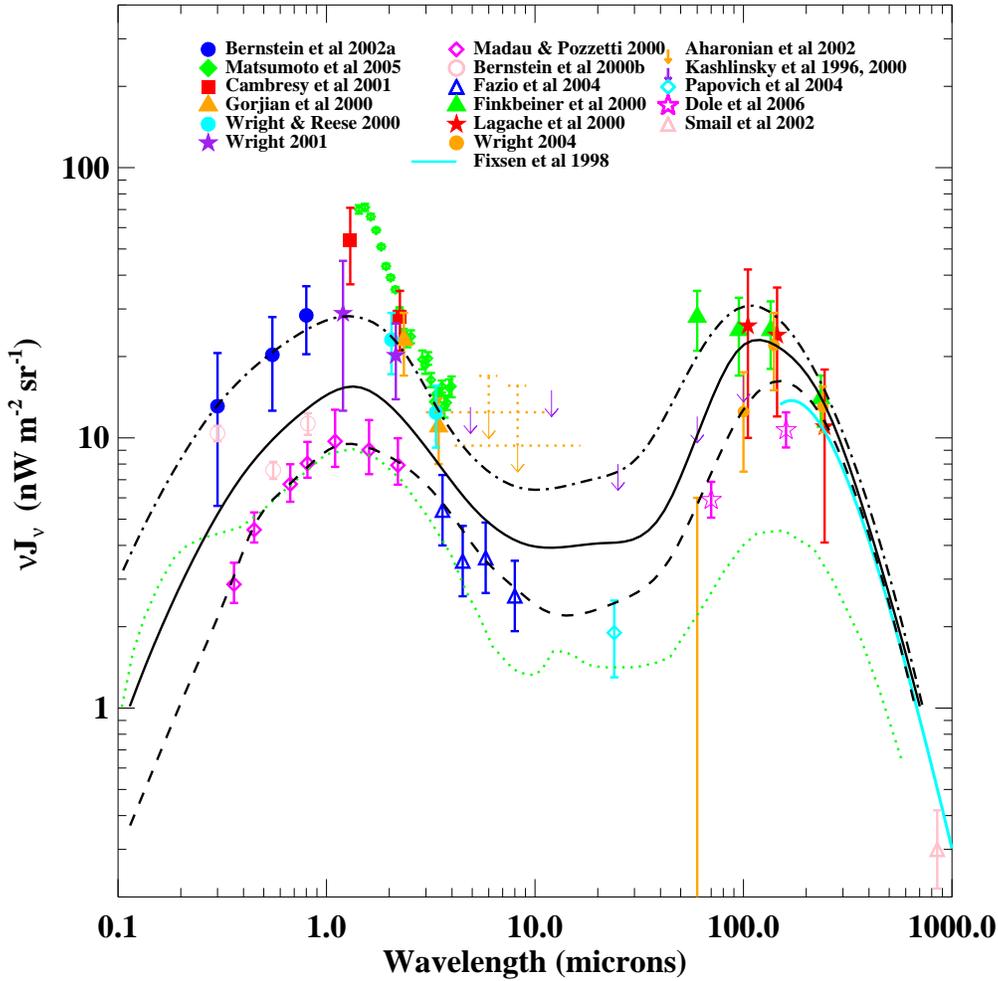}
\caption{
\label{fig.obsbackgd}
The observed extragalactic background light.  Sources for
observational points and limits are given in the legend.  Filled
symbols represent absolute measurements of the background light, while
empty symbols represent integrated galaxy counts.  The filled circles
for the Bernstein \ea (2002a,b) absolute background incorporate the
changes in their erratum.  We show the analytic fit to the FIR
measurement of \citet{fixsen98}.  For the gamma-ray limits of
\citet{aharonian02} we show squares corresponding to their tentative
detections of $\tau=1.5$ at 5.5 TeV for Mkn 501 and $\tau=5.5$ at 4.0
TeV for H1426+428. We convert these limits to effective IR wavelength
and background intensities using the approximations of
\citet{aharonian01}.  Upper limits from
\citet{kashlinsky96a,kashlinsky96b} and \citet{kashlinsky00} use the power
spectrum to estimate the total extragalactic contribution.  The
dashed, solid, and dot-dashed black curves show the traces through
this data that are discussed in the text.  The dotted curve shows
results from the semi-analytic model of \citet{primack05}.}
\end{figure*}
}

\newcommand{\madaufig}{
\begin{figure*}
\includegraphics[width=14cm]{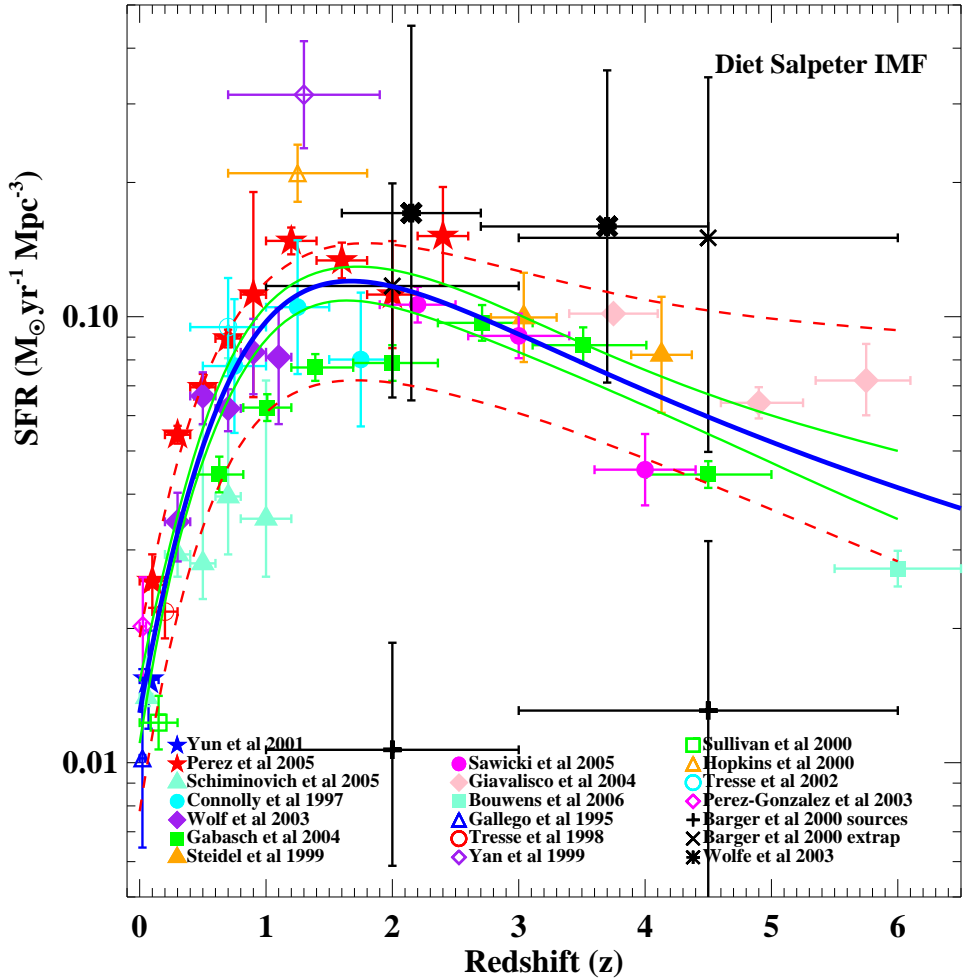}
\caption{
\label{fig.madau}
Star formation rate per unit comoving volume as a function of redshift
as derived from observations.  The observational points,
which are labelled by source, have been rescaled to our chosen ``737''
cosmology, and to a common set of luminosity and completeness
calibrations using the ``diet'' Salpeter IMF as discussed in the text.
We use uniform dust correction factors for the optical and UV points
(see the text for details). The thick solid line is our best fit to the data.
The thin solid and thin dashed curves indicate the $1\sigma$ envelope
of the fits in our standard and ``random-calibration-error'' 
samples of fits, respectively.  }
\end{figure*}
}

\newcommand{\rhoderivfig}{
\begin{figure*}
\includegraphics[width=14cm]{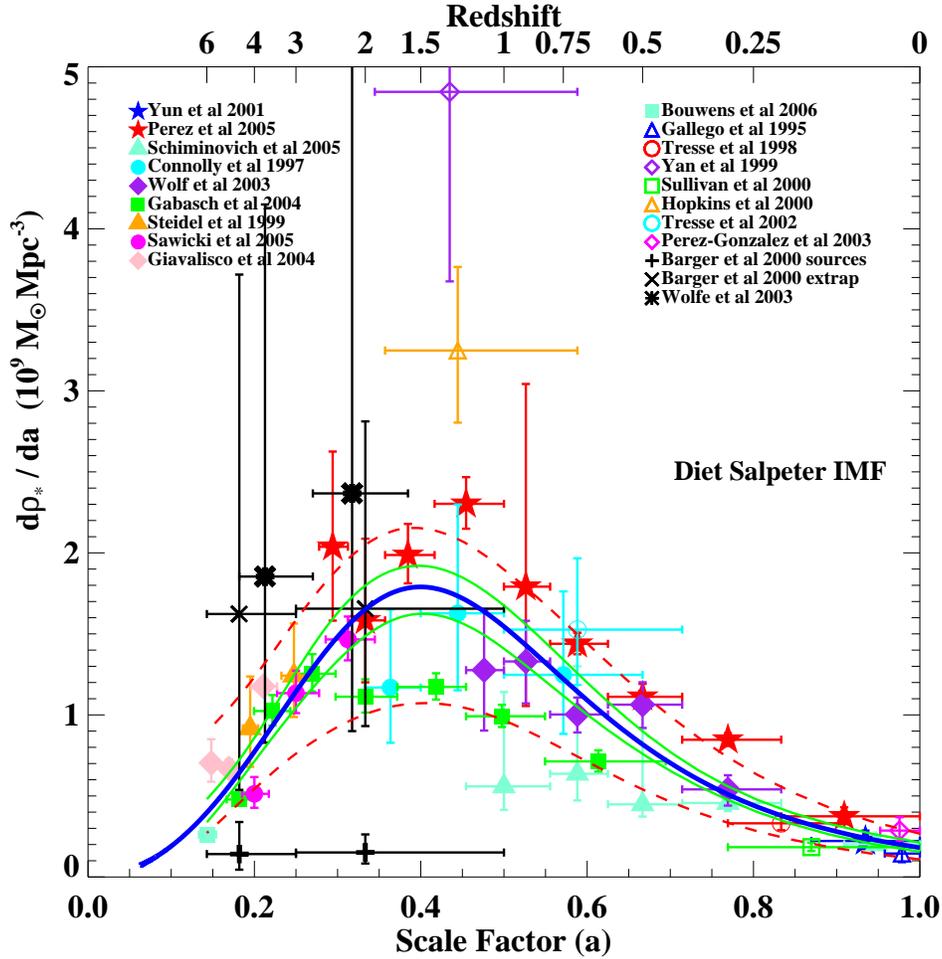}
\caption{
\label{fig.rhoderiv}
Star formation rate per unit scale factor versus scale factor.
This is similar to Figure \ref{fig.madau}
but with rescaled axes.  The vertical axis here is the
derivative of the mass of stars formed with respect to the scale
factor, $d\rhoform/da = (1+z) \rhodotform / H(z)$.  In this form the
data points are independent of cosmology, and the total star formation
is the area under the data points.  One can easily see the relative importance
of different redshift ranges.  }
\end{figure*}
}

\newcommand{\starmasszfig}{
\begin{figure*}
\includegraphics[width=14cm]{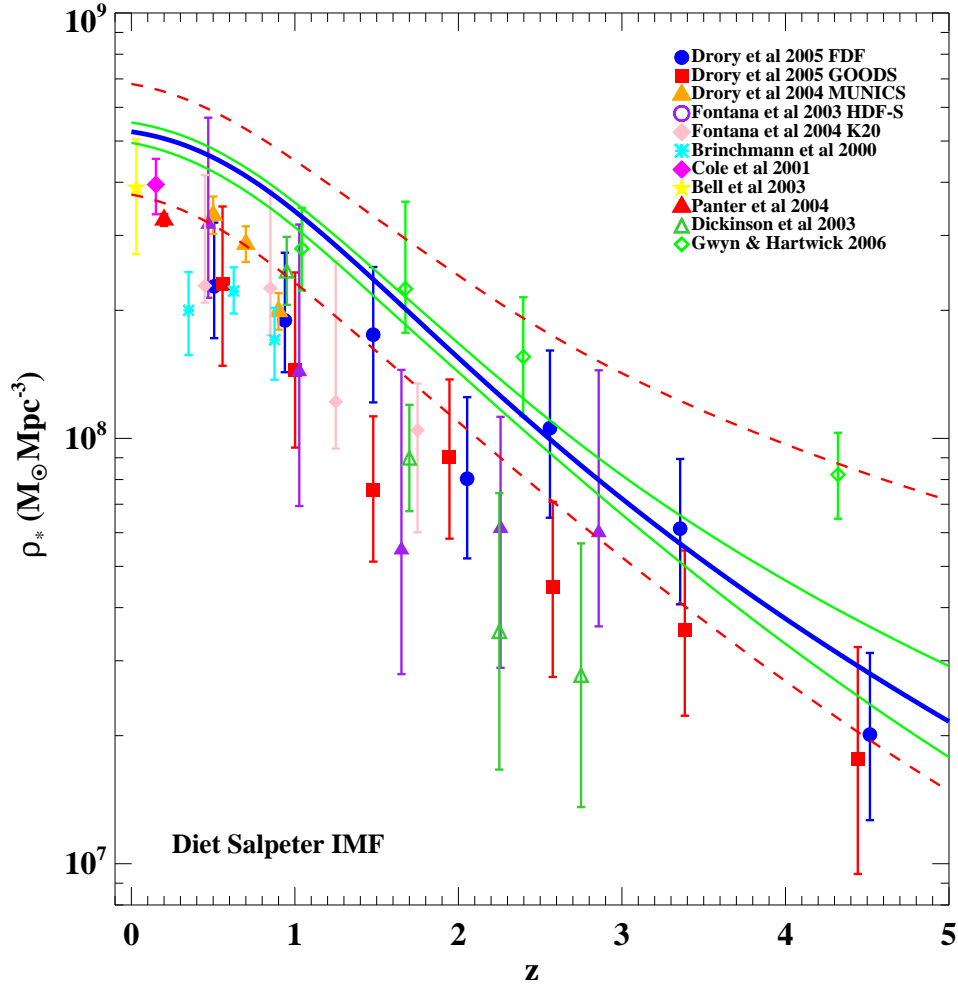}
\caption{
\label{fig.starmassz}
Stellar mass as a function of $z$ scaled to a diet Salpeter IMF (at $z=0$
this scaling factor is 0.70 and it slowly increases to 0.74 at $z=5$).
Sources of the data points are as indicated.
The heavy solid line shows the result of integrating our best fit through 
Figure \ref{fig.rhoderiv}, including the appropriate recycling factor.
The light solid and dashed lines show the $1\sigma$ boundaries of our
``normal'' and ``randomly-recalibrated'' ensembles of fits,
respectively.
}
\end{figure*}
}

\newcommand{\kbgcontours}{
\begin{figure}
\includegraphics[width=7.3cm]{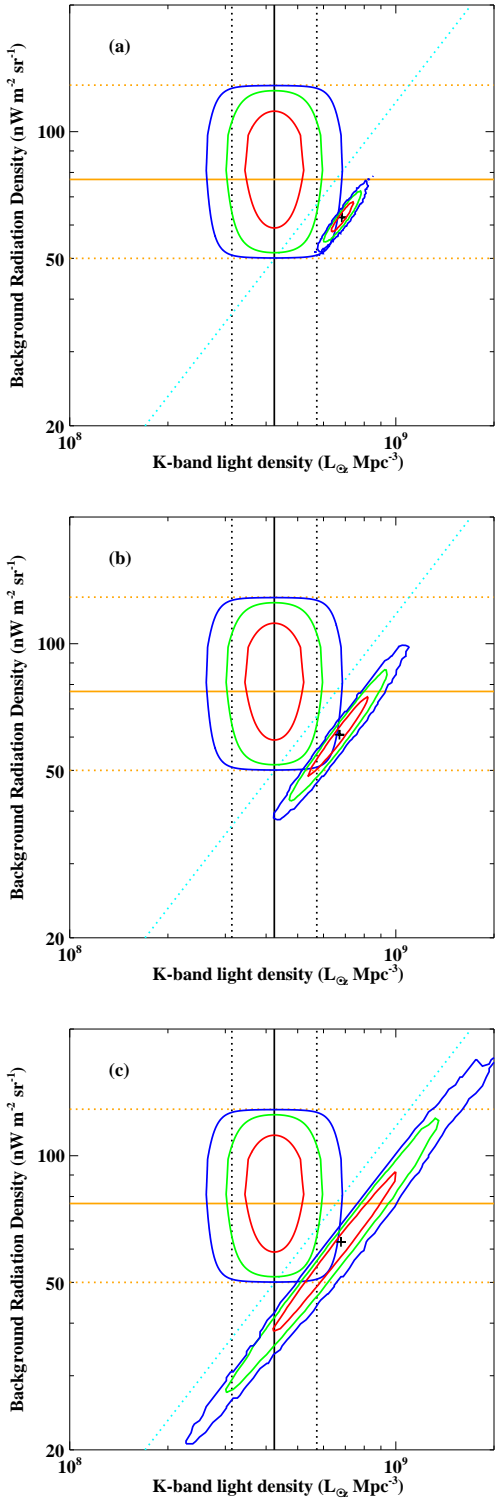}
\caption{
\label{fig.kcontours}
The combined range of $K$-band light and EBL
allowed by the range of SFHs.   In each panel,
the horizontal
lines denote the minimum, best guess, and maximum background intensities.
The vertical lines denote the best fit and $2\sigma$ error bars
on the $K$-band luminosity density.  
The vertically extended rounded 
contours show the joint 67\%, 95\%, and 99.8\% joint confidence intervals on
these two observed parameters.  The diagonally extended contours show
the relation of the two quantities in the ensemble of SFH fits,
for the following assumptions:
(a) No freedom in SFR calibration.
(b) Freedom of $\sigma_A = 0.2$ in SFR calibration
(see explanation in text).
(c) Freedom of $\sigma_A = 0.5$ in SFR calibration.
}
\end{figure}
}

\newcommand{\mbgcontours}{
\begin{figure}
\includegraphics[width=7.5cm]{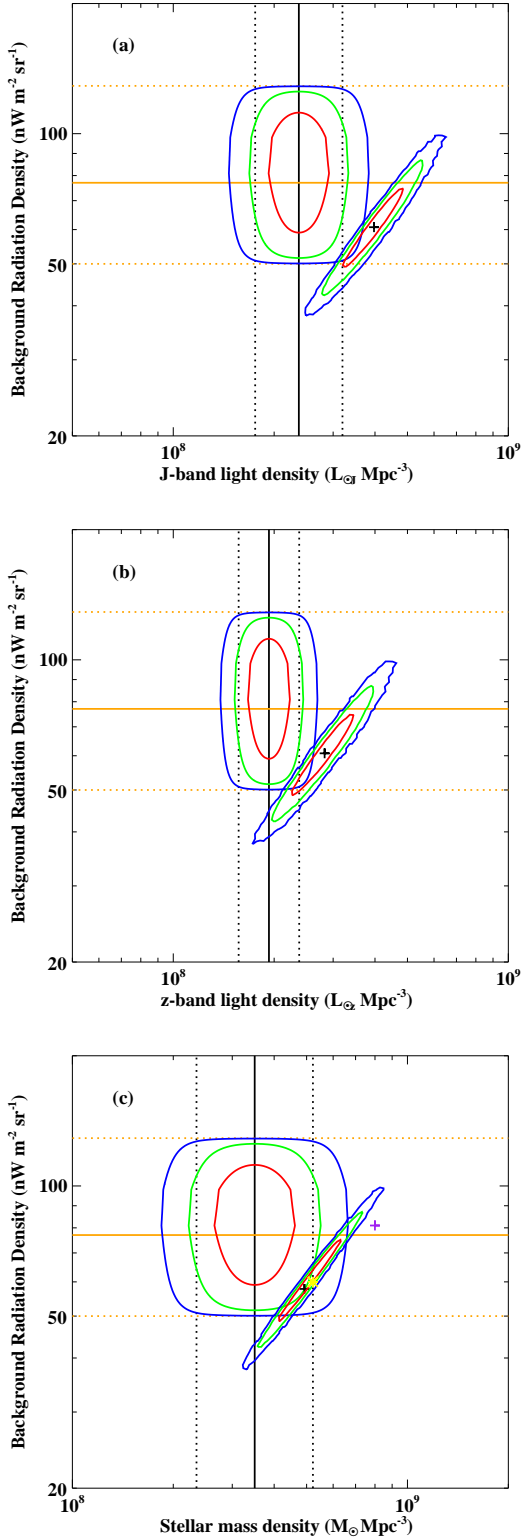}
\caption{
\label{fig.masscontours}
The contours of EBL vs.\ 
(a) $J$-band stellar light density,
(b) $z$-band stellar light density, and 
(c) stellar mass density 
allowed by the range of SFHs.  Except for
substituting these quantities for $K$-band light, 
the plots are the same as in the previous figure.  In panel (c),
the asterisk shows the estimated relation of stellar mass and 
EBL from \citet{madau01}, rescaled
to our IMF, and the cross shows the result from the simulation of
\citet{nagamine00} (see text for details).  }
\end{figure}
}

\newcommand{\imfplot}{
\begin{figure*}
\includegraphics[width=16cm]{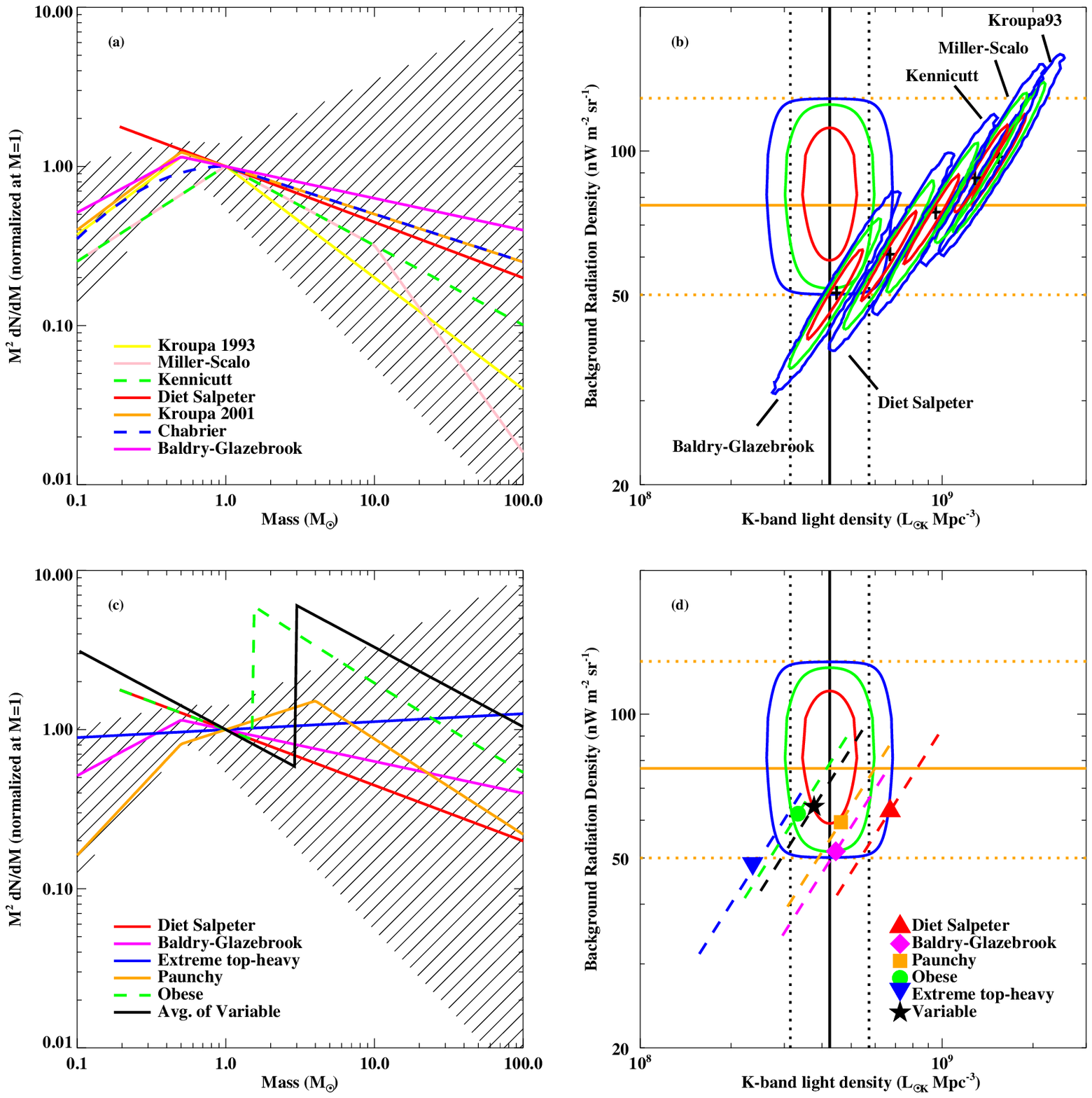}
\caption{
\label{fig.imfplot}
(a) 
The standard IMFs considered in the paper.  The vertical axis is
the mass contained in the IMF per unit logarithmic stellar mass.
The shaded region is the allowed region taken from \citet{kroupa02}.
(b) 
The same as Figure~\ref{fig.kcontours} but showing the 
effects of using the different IMFs in the previous panel.
We scale each SFH to account for
the UV luminosity in a $10^8 \yr$ burst for that particular IMF.
The contours reflect a systematic error of
$\sigma_A = 0.2$ in the SFH calibration.  
(c) 
Shows the new IMFs introduced in the paper.
We also plot Baldry-Glazebrook as an example of a moderately top-heavy
IMF and Diet Salpeter for comparison.
(d) 
Like panel (b) but for the IMFs shown in 
panel (c).
To avoid the confusion of overlapping contours, we
only show the values from the best fit and indicate the 
approximate contour extent by a dashed line.
}
\end{figure*}
}

\section{INTRODUCTION}

In the last decade, observations have built up a picture of when and
where stars form in the universe.  From inventories of the
stellar content of the local universe \citep{cole01,bell03}
and surveys of star formation over its 
entire cosmic history \citep[e.g.][]{madau96,gabasch04}, 
it appears that star formation was
much more rapid in the past, reaching a gentle peak at $z \tsim 1$--3
and falling off towards higher redshifts.  These general trends are in
accord with the $\Lambda$-dominated cold dark matter (LCDM) model of
structure formation that has become widely accepted in recent years.
This model successfully explains the spectrum of perturbations in the
microwave background, the clustering properties of local galaxies, the
optical depth and fluctuation spectrum of the Ly$\alpha$ forest gas,
and the flux from distant supernovae \citep{spergel06}.  
However, galaxy formation is
less well understood than these other phenomena.  Theoretical models
of galaxy formation can be implemented either by using semi-analytic
calculations \citep[e.g.][]{kauffmann99,cole00,somerville01}
or by direct numerical simulations \citep[e.g.][]{katz92,springel03};
each method has its
own well-known advantages and drawbacks.  
Difficult issues that affect
the results in both methods include galaxy supernova and AGN
feedback \citep{springel05,dekel06}, 
the history and physical influence of reionisation \citep{gnedin00,bullock00}, 
the influence of ``cold mode'' accretion \citep{keres05}, 
and the properties of zero-metallicity stars \citep{tumlinson04}.
Different assumptions about these issues can produce quite different histories
of star formation, so figuring out the total amount of star formation
is a vital clue to the physics.

The observational measures of star formation are sufficiently
imprecise that cross-checks using different measures are important.
The star formation history (SFH) is measured by the direct emission from
young stars or reprocessed radiation from dusty star-forming regions.
This can be tested against the evolution of the
stellar mass, using light from old stars (preferably in the near-infrared) as
a proxy for mass; this has been carried out by many authors (e.g.,
\citealp{cole01,dickinson03,gwyn05,hopkins06}).  The integrated star
formation rate tends to exceed the observed stellar mass, especially
at high redshift.  Given the large statistical and calibration errors
in both quantities, however, the evidence for a discrepancy is not
very strong.  The observed metal density can be used as well
\citep{madau00, hopkins06}, though the results from this approach are not yet
precise enough to indicate more than a very approximate agreement.
\citet{hopkins06} introduced the use of neutrinos from Type II
supernovae to place an upper limit on the normalisation of the star
formation history. 
Uncertainties in the effective neutrino temperature and stellar mass
thresholds of Type II SNe cause systematic uncertainties in this limit;
with present data, it is probably best to view the limit as constraining
the supernova physics given the estimated SFH.

The extragalactic background light constitutes another vital test of
the star formation history, since it records the total stellar
emission over all time, weighted by the scale factor at the time of
emission.  The observed background light contains two humps, one in
the far infrared (FIR) and one in the ultraviolet (UV) to
near-infrared (NIR) range.  The FIR hump primarily owes to thermal
dust emission from heavily obscured, rapidly star-forming galaxies.
There is much debate over the optical/NIR portion, in which the sum of
the observed galaxies amounts to $\lesssim 50$\% of the total detected
background \citep{wright01,bernstein02a}.  The observed spectrum in
this region might contain a hint of Lyman-$\alpha$ emission from
low-metallicity, high-mass Pop III stars at $z>6$ \citep{matsumoto05},
but observational uncertainties and theoretical difficulties make this
idea very controversial.  It is certainly possible that most of the
excess optical/NIR background light comes from the outer regions of
normal galaxies or stars in faint undetected galaxies
\citep{bernstein02c}.  

Many authors have attempted to tie the background radiation in
different wavebands to the history of specific subsets of galaxies,
e.g. FIR emission from dusty starbursts or the NIR emission from
pre-reionisation epoch galaxies.  \citet{madau00} were the first to
consider the {\em total} amount of the background radiation as a
record of the cumulative stellar emission from all types of galaxies.
This exercise is valuable because reprocessing of stellar radiation by
dust and atomic continuum opacity, which can be very difficult to
model correctly, will in general not significantly change the total
amount of emitted energy.  The stellar background energy is dominated
by the stars that have already evolved off the main sequence, down as
low as $0.9 \msun$, so it samples a broader range of stellar masses than
the direct indicators of the SFH.  Madau \& Pozzetti (2000; see also
\citealp{madau01}, \citealp{pozzetti01}) found good agreement between
the background radiation and the total amount of stars in the
universe, concluding that there was little room for alternative
sources of energy such as an early generation of stars with a
top-heavy IMF.  Since the work of Madau \& Pozzetti, additional
measurements of the background have been made in several different
bands, and our understanding of the star formation history and local
stellar density has improved.  It thus seems worthwhile to test the
background radiation as an indicator of the star formation history
once more.

In \S\ref{sec.calib}, we discuss how we calibrate various indicators
of the stellar mass and the star formation rate, and discuss the range
of masses contributing to each.  In \S\ref{sec.obs}, we discuss the
observed cosmological background radiation, star formation history,
and stellar mass density in the universe and we construct an ensemble
of fits to the SFH to test the sensitivity of other observables to it.
In \S\ref{sec.compare}, we compare the results of these various
indicators of star formation to one another.  Using the ensemble of
star formation histories, we conclude that it is difficult to
reconcile the background radiation with the stellar mass density
observed today, using our default IMF assumptions.
\S\ref{sec.discuss} discusses several possible solutions to this
problem, focusing on the role of the stellar IMF.  We show that the
observed ratio of the total background light to the present-day NIR
luminosity density is a powerful test of the IMF.  Despite the large
observational uncertainties, we can significantly restrict the allowed
set of IMFs.  \S\ref{sec.conclude} summarises our conclusions.

Since the IMF figures in every stage of our analysis and emerges
as a focal point of our discussion, it is worth making a few 
introductory comments about its role.  Indicators of the instantaneous
star formation rate are dominated by massive stars;
in our standard calculations below, for example, half of the UV
luminosity comes from stars with mass $M \gtrsim 15 \msun$.
The contribution of a stellar population to the extragalactic
background depends on the integrated bolometric luminosity over
its lifetime.  This includes a significant contribution from less
massive stars, and in our standard calculations half of the
background comes from stars with mass $M>3M_\odot$.
The present $K$-band light density comes largely from 
sub-giants and giants with masses just above the main-sequence
turnoff, typically $M\sim 1M_\odot$ for an old stellar population.
The expected ratio of these three measures of the cosmic
star formation history, therefore, depends on the shape of the
IMF above $1M_\odot$.  Much of the {\it mass} in the IMF
resides in low mass stars, and there has been a great deal
of observational investigation of the IMF shape below $1M_\odot$.
However, low mass stars emit very little light, so IMF
changes in this regime have almost no impact on the {\it relative}
values of these three probes of the cosmic star formation history;
they change the expected/inferred stellar mass density for the
three measures by the same factor.
Our conclusions in this paper are, therefore, almost entirely
insensitive to the adopted shape of the IMF below $\sim 0.8 M_\odot$.

Throughout this paper, we assume a flat ``737'' LCDM cosmology of 
$\Omega_M = 0.3$, $\Omega_\Lambda = 0.7$, 
$h \equiv H_0 / (100 \kms \Mpc^{-1}) = 0.7$, and
$\Omega_b = 0.02 \, h^{-2}$, consistent with values derived
from the {\it WMAP} experiment \citep{spergel06}.
\section{CALIBRATIONS OF STELLAR ENERGY OUTPUT}

\label{sec.calib}

\imftable

In the central parts of some galaxies, the mass in stars can be
estimated directly from its dynamical effects, but on larger scales
the dynamics are dominated by dark matter.  Hence, at present, all direct
estimates of the global production of stars are based on the light
they produce.  The calibration of the mass-to-light ratio is sensitive
to the age and metallicity of the population, but these can be
constrained using the observed colours and/or spectral lines.  The
other major uncertainty is the stellar initial mass function (IMF).
The most conventional choice is the original Salpeter IMF, with a
mass range 0.1--$100 \msun$.  However, there is evidence that this
contains too many low-mass stars, as shown both by local field star
observations \citep{gould96} and by estimates of the dynamical mass in
ellipticals \citep{bell01}.  The latter authors advocate a ``diet''
Salpeter IMF, which they create by truncating the IMF at the low end
so that it contains 0.7 times the mass in the {\em
  current} stellar population, when normalised to the same high-end
amplitude.  Using a representative estimate of the cosmic star
formation history from our best fit below, we find that this
implies 0.788 times as much mass in the IMF before the high-mass stars
burn away; the indicators of star formation rate per mass of stars
formed are thus shifted by the inverse of this factor.  (This scaling
is not very sensitive to the history.)  These two IMFs and some other
forms that we consider later in this paper are summarised in
Table~\ref{table.imf}.

To convert between the star formation history and observed measures
such as the stellar mass, UV continuum, and $K$-band light, we use the
PEGASE.2 stellar population code
\citep{fioc97}.\footnote{http://www.iap.fr/users/fioc/PEGASE.html}
This code contains a large number of options for the stellar physics.
Our conversions assume a close binary mass fraction of 0.05, evolutionary
tracks with stellar winds, the SNII model B of Woosley and Weaver (1995),
and a constant metallicity of $Z=0.02$.  To assess the sensitivity of our
results to this particular
stellar population code, we have compared them to those
obtained using the GALEXEV-2003 code of Bruzual \& Charlot, and find only
small differences of $\tsim 10$\% at most (see also Bruzual \& Charlot
2003, where some additional comparisons between the codes are presented).

\sfrconvtable

When constructing the cosmic SFH one needs to convert from observed
luminosity to star formation rate (SFR).
Table~\ref{table.sfrconv} lists the conversion factors
of different star formation indicators for three different IMFs.  
The conversion factors are $f_{bol} \equiv L_{bol} / \sfr$,
$f_{H\alpha} \equiv L_{H\alpha} / \sfr$, and 
$f_{UV} \equiv L_\nu / \sfr$.  
The latter quantity is nearly constant over the wavelength range
1500--2800 \AA\ used in various UV galaxy surveys, since
$L_\nu$ itself is nearly constant before extinction 
for burst lengths of $\tsim 10^8$--$10^9 \yr$;
hence we choose to evaluate $f_{UV}$
at 2000 \AA\ without loss of generality.  We calculate these constants with
PEGASE, using constant-SFR bursts of various duration.  For a given SFR, the
observed luminosity depends on how long that SFR rate persisted before the
observation.  Since surveys are flux limited they
are biased towards observing galaxies undergoing rapid
bursts, boosting their luminosities. Therefore,
one usually assumes a short duration
burst when making the conversion to SFR.  In Table~\ref{table.sfrconv} we
show the conversion factor for three different assumed burst lengths of $10^7$,
$10^8$, and $10^9$ years. 
As shown in the table, this uncertainty in burst duration introduces
uncertainties in the derived SFRs of up to a factor of two.
In our plots we assume burst lengths of $10^8$ years.
We note that assuming such globally short burst lengths cannot be valid when
integrating over the entire galaxy population, as is done to get the
total SFR at any epoch, since this includes galaxies that were too faint to be
included in the survey and hence had no bias to be observed while bursting.
However, correcting for this bias with a
fully self-consistent approach is beyond the scope of this paper.

Furthermore, the UV flux depends somewhat on metallicity.  Using PEGASE we find
that for metallicities of $Z=0.002$ and $Z=0.008$ the UV output per 
unit of star formation
is roughly 25\% and 10\% larger, respectively, than for the solar
value of $Z=0.02$, for Gigayear bursts.  However, \citet{panter03}
found that the mean metallicity of star-forming gas has been close to
solar for $\tsim 6 \Gyr$ into the past, suggesting there is not much
of an offset in the average UV calibration for much of the Universe's
history.  We estimate that metallicity effects can introduce systematic
calibration uncertainties of $\sim 10\%$, but we will not attempt a more
detailed treatment until the cosmic metallicity history becomes better
known.  Finally, a change in the upper cutoff mass, from 100 to $150 \msun$, for
example, would change the bolometric and UV output by only a few
percent.  The SFR tracers are produced mostly by stars with $M >
5$--$30 \msun$, depending on which indicator is used.

Another required quantity for our study is the mass remaining in a stellar
population, after the higher-mass stars burn away.  We use PEGASE to calculate a
table of total stellar mass versus time after a burst of star
formation.  We include in this total white dwarfs and neutron stars,
which at late times can be a marginally significant fraction of the
mass.  We then convolve this table by an assumed cosmic star
formation history $\rhodotform(z)$ to get the remaining present-day
stellar mass density, i.e.
\begin{equation}
  \rhorem = 
  \bar{f}_{mass} \, \rhoform = 
    \int_0^{t_0} \rhodotform(z) \, f_{mass}\left[t_0 - t(z)\right] \, dt
  \; .
  \label{eqn.rhostar}
\end{equation}
The ``mass-lockup'' fraction $\bar{f}_{mass} \equiv \rhorem /
\rhoform$ is very sensitive to the low-mass form of the IMF and only
slightly sensitive to the star formation history, for reasonable
choices of both.  For our standard star formation history derived
below, we find lockup fractions of $\bar{f}_{mass} = 0.71$ and 0.63
for Salpeter and diet Salpeter IMFs respectively.  This mass is
dominated by main-sequence
stars with $M \sim 0.5 \msun$ in the latter case.  
However, as noted in the introduction, the low-mass form of the IMF
sets the calibration between any of our luminosity measures and
the total stellar mass, but it has almost no effect on the relative
values of the three measures, and therefore on our eventual conclusions.

The $K$-band light produced by the current stellar population is
given by a similar equation,
\begin{equation}
  \rho_K = 
    \int_0^{t_0} \rhodotform(z) \, l_K\left[t_0 - t(z)\right] \, dt
  \; .
  \label{eqn.klight}
\end{equation}
Here $l_K$ is the $K$-band luminosity per unit stellar mass
of a stellar population as a function
of its age.  For a specified IMF, $\rhorem$ and $\rho_K$ are
very tightly correlated since they are both dominated by low-mass
stars.  The $K$-band luminosity is dominated by red giant branch (RGB)
and asymptotic giant branch (AGB) stars with $M \approx 1.0 \msun$.
At a lower metallicity of $Z=0.002$, we find $\rho_K$ is lower by $\tsim 5\%$
for plausible star formation histories, because of the bluer colour of 
low-metallicity giants.  

We obtain results in $J$ and $z$ bands by a similar method.  $\rho_K$
as given by Equation~\ref{eqn.klight} is the unextincted light, but
even in $K$-band there is a small amount of extinction.  For an
extinction estimate, we adopt the ``fiducial'' dust model from the
global luminosity density modelling of \citet{baldry03}.  This model
implies effective optical depths in the $K$, $J$, and $z$ bands of 
$\tau_K = 0.08$, $\tau_J = 0.15$, and $\tau_z = 0.22$.
Some idea of the uncertainty in these values can be obtained from the
three models in Figure~6 of Baldry \& Glazebrook, along with their
fiducial model; the dispersion in the values from these four models is
0.03, 0.05, and 0.07 for the $K$, $J$, and $z$ bands, respectively.

Given the star formation history of the universe $\dot{\rho}_\star(z)$, we
can easily compute the resulting bolometric background, which we will refer
to interchangeably as the extragalactic background light (EBL).
The background is given by an integral over cosmic time:
\begin{equation}
  \jbg = \frac{c}{4\pi} \int_0^{t_0} \rhodotform(z) \, S_{\it EBL}(z) \, dt
   \equiv \frac{c}{4\pi} \rhoform  \bar{S}_{\it EBL}
  \; .
  \label{eqn.backgd}
\end{equation}
Here the background contribution function $S_{\it EBL}(z)$ gives
the total background light energy per unit mass
formed into stars at a redshift $z$:
\begin{equation}
  S_{\it EBL}(z) = \int_{t(z)}^{t_0} 
     \frac{l_{\it bol} \left[t_{em} - t(z) \right]}{1+z(t_{em}) } \, dt_{em}
  \; .
  \label{eqn.impulse}
\end{equation}

We use PEGASE to calculate the bolometric luminosity per unit mass
$l_{\it bol}(t)$ for a given choice of IMF.  For stars with lifetimes
much shorter than the lookback time $t_0 - t(z)$, the contribution to
the integral is simply $E_{bol}(M) (1+z)^{-1}$, where $E_{bol}(M)$ is
the total energy released from nucleosynthesis in a star of initial
mass $M$ over the lifetime of the star.
For less massive stars, the contribution is enhanced by the larger
scale factor $[1+z(t_{em})]^{-1}$,
especially during the more luminous, post-main-sequence phase 
at the end of the star's life.  At still
lower masses, where the stellar lifetime exceeds the lookback time and
the stars do not have time to release most of their energy, the
contribution then diminishes rapidly.  Over the entire range of
stellar masses, $E_{bol}(M)$ is roughly proportional to the mass.
Hence, 
{\it the contribution to the background roughly tracks 
the integrated mass in the IMF above the main sequence turnoff.}
For an IMF of Salpeter
slope, about half of the total energy that the stellar population will
give up within the present-day age of the universe is emitted within
the first $0.13 \Gyr$, or above a turnoff mass of $4.6 \Msun$.  The
background light is biased to slightly lower masses; for a realistic
SFH, half of the background comes from stars above $\tsim 3 \msun$.

The background contribution function $S_{\it EBL}(z)$ is maximised at
a lookback time of about 2.3 Gyr or a redshift of 0.2, for a Salpeter or
diet Salpeter IMF with a peak value of
$1.0 \times 10^{10} \Lsun \msun^{-1} \yr$ 
for a diet Salpeter IMF.  Despite the $(1+z)^{-1}$ dimming, $S_{\it EBL}$
from high redshift star formation remains above 0.3 times its maximum value
owing to contributions from long lived, lower mass stars.
(\citealp{madau00} contains some plots of $S_{\it EBL}(z)$ for Salpeter-like
and top-heavy IMFs.)
We note that Bruzual \& Charlot models give very similar results
for the bolometric luminosity; specifically they are larger by about 4\% on
average, which is well within our errors.
The derived bolometric background can also be affected by metallicity; it goes
up by $\tsim 20\%$ when the metallicity is decreased from solar to 
$Z=0.002$, assuming plausible SFHs,
though as noted above the average metallicity is roughly solar
many Gigayears into the past.

\section{OBSERVATIONAL MEASURES}
\label{sec.obs}

\subsection{Total background radiation}
\label{sec.obs.backgd}

Figure \ref{fig.obsbackgd} shows measurements of the EBL as a function
of wavelength.  This plot shows absolute measurements of the EBL
(solid symbols), and estimates resulting from the integrated light of
galaxies (open symbols).  The latter are lower limits to the
background flux, since faint or low-surface-brightness galaxies or
exotic (non-stellar) sources of energy may also contribute to the
background light.  The EBL is dominated by two main peaks in the
optical and infrared, corresponding to direct and dust emission.  For
now let us assume that all of this energy ultimately derives from the
radiation of stars.  The units we use for both the differential ($\nu
J_\nu$) and total ($\jbg$) background radiation are ``bgu''$\equiv \mbox{nW} \,
\mbox{m}^{-2} \sr^{-1} = 10^{-6} \bgunit$.  A comprehensive review of
the background is given in \citet{hauser01review}.

\obsbackgdfig

Integrated galaxy counts in the optical/NIR region were compiled by
\citet{madau00}, who suggested that the flux had essentially
converged.  In contrast, \citet{bernstein02c} argued that the
isophotal magnitudes used in these estimates failed to capture much of
the flux from the faint portions of galaxies, and found substantially
higher estimates from an ``ensemble photometry'' technique, as shown
by the open circles in the figure.  \citet{fazio04} provide integrated
galaxy counts in the NIR from Spitzer, which are not as close to
convergence as those in the optical.

Direct measurements of the absolute background by 
Bernstein, Freedman, \& Madore (2002a,b)
give substantially higher values.  These authors attribute the
differences in the optical region to two effects: flux missed by
normal galaxy count surveys in the outer, diffuse parts of galaxies
(partially recovered by the ensemble photometry technique) and
high-redshift galaxies that are missed altogether by these surveys
owing to their low surface brightness.  Direct measurements in the
NIR, using the space instruments DIRBE
\citep{cambresy01,gorjian00,wright00,wright01} and IRTS
\citep{matsumoto05}, mainly give even higher estimates, particularly
the latter.  The dispersion among these DIRBE points is mainly
from different ways of analysing the foregrounds, including different
ways of subtracting stars and different models for the zodiacal light.
The zodiacal light subtraction is likely the largest source of
systematic error in these direct background measurements, especially
since the shape of the NIR excess nearly matches that of the zodiacal
light \citep{dwek05}.  It is perhaps encouraging that all of the
optical and NIR direct background estimates from different groups give
values significantly higher than the direct counts, suggesting that at
least part of the excess signal is real.  To many authors, the
apparently peaked NIR excess suggests a Ly$\alpha$-dominated
contribution from Pop III stars before reionisation at $z > 10$ (e.g.,
\citealp{santos02}).  However, this model has some major problems
stemming from its large energy requirements \citep{madau05} and
observational constraints on the number of NIR-detected Ly$\alpha$ 
emitters \citep{salvaterra06}.

Turning to the mid- and far-IR region,
we show integrated galaxy counts derived from Spitzer observations at
24 \micron\ \citep{papovich04}.  We also show the integrated light
derived from an analysis at 70 and 160 \micron\ by \citet{dole06}, who
stacked the observed 24 \micron\ sources and integrated the
resulting light in these other bands.  In both cases, we show only the
flux resulting from the observed counts, rather than their
extrapolation of the counts to zero flux, which
would raise the plotted points by 20--40\%.  The faint counts in these
analyses are heavily dependent on estimating the detection
completeness against the background of unresolved sources, and future
work on this issue may well result in slightly different faint-end
slopes, which would strongly affect the extrapolation.

The absolute background detections in the far-infrared (FIR) region
come from the DIRBE and FIRAS experiments on board COBE.  For the
DIRBE data originally obtained by \citet{hauser98}, different groups
have treated the photometric calibration and foreground subtraction in
different ways, leading to multiple results from the same data
\citep{finkbeiner00,lagache00,wright04}.  At 240 \micron\, the
different results are nearly in agreement, but as the wavelength
decreases the scatter among them increases, probably because of the
greater complications associated with the higher zodiacal light
intensity.  These shorter-wavelength points can have a strong effect
on the estimated height and width of the FIR peak.  We have
represented the longer-wavelength FIRAS data by the analytic fit in
\citet{fixsen98}, rather than the noisier raw data.

TeV $\gamma$-ray measurements towards blazars can in principle measure
the background via the opacity to 2-photon scattering.  In our
judgement, these measurements are not yet very reliable; the estimated
optical depth is extremely sensitive to the assumed shapes of both the
intrinsic TeV source and the background spectrum itself.  Hence at
this point, we prefer to use direct absolute background detections,
even though they are uncertain.  However, in the MIR range only loose
upper limits exist as the galaxy signal is too low to compete with the
zodiacal light \citep{hauser98}.  We thus show results of optical
depth models towards two blazars presented in \citet{aharonian02},
using the formulae of \citet{aharonian01} to convert optical depth to
MIR background intensity.  These arrows are tentative detections but
should probably be treated as upper limits.  We note that TeV
$\gamma$-ray absorption estimates should be best at constraining
sharp, intense features in the background; it would be at least
somewhat surprising if the NIR peak shown by the IRTS data is
consistent with existing TeV data \citep{aharonian05}.  To strengthen
the case for a lower EBL in the MIR region, we show the results of
fluctuation analyses by \citet{kashlinsky96a}, \citet{kashlinsky96b},
and \citet{kashlinsky00}, which attempt to place limits on the galaxy
contribution from the power spectrum of the background rather than its
absolute intensity.

As mentioned above, and as is apparent in the galaxy counts in
Figure~\ref{fig.obsbackgd}, the galaxy contribution to the
extragalactic background is expected to have two broad humps.  A model
of the background radiation \citep{primack05} is shown by the dotted
line; while this clearly underestimates the background intensity
especially in the FIR, it indicates the expected shapes.
For our work, we require better estimates of the total
intensity.  We show three traces through this plot, representing what
we consider the minimum (dashed line), maximum (dot-dashed line), and
best guess (solid line).  These curves are simply constructed to be
smooth curves without a particular functional form, but their shapes
do reflect global galaxy emission models.  The ``max'' trace uses the
background measurements in the optical and NIR and the MIR gamma-ray
limits, while the ``min'' trace follows the lower values implied by
the galaxy counts.  The ``mid'' trace adopts a compromise between the
higher galaxy counts and the lower range of absolute background
estimates.  In the FIR, the
differences between these curves are unrelated to those in the
optical, and mainly reflect the assumed width of the FIR peak.  The
total background fluxes for these traces are 50, 77, and 129 bgu.  If
we consider only the optical/NIR portion below 10~\micron, we get 22,
36, and 70 bgu, while if we consider only the FIR portion above 10
\micron, we get 28, 41, and 59 bgu.

Our EBL estimates range to substantially higher values than the best
estimate of $55 \pm 20$ bgu determined by \citet{madau00}, which was later
revised and increased to 60 bgu by \citet{madau01}.  Later
estimates, which include 60--93 bgu by \citet{gispert00}, 45--170 bgu
with a preferred value of 100 by \citet{hauser01review}, and $100 \pm
20$ bgu by \citet{bernstein02c}, are more in line with our estimate.
The crucial issues contributing to this uncertainty are not
statistical error but systematics like the calibration of the FIR
measurements, the zodiacal light subtraction in the optical and NIR
background measurements, the treatment of the falloff from the optical
and FIR peaks to the MIR region, and simply which data points to
include.  Illustrating the latter issue,
\citet{dole06} recently discussed the EBL in the context of their FIR galaxy
stacking analysis.  
They rely on limits from TeV $\gamma$-ray
absorption and IR fluctuation analyses to set the upper limits on the
flux, whereas we have preferred absolute background detections.  The
upper and lower boundaries of their allowed region are shown in their
Figure 12.  Using these traces and adopting a higher-order integration
scheme than used in their derivation, we find that the $\jbg$ is in the 
range of 52--77 bgu, spanning approximately the lower half of the allowed 
range we find here.

Below, we will need to estimate the probability (or likelihood)
distribution of $\jbg$.  It is clear that this distribution is far
from Gaussian, as the galaxy counts provide a hard lower limit to the
background in certain portions of the spectrum, but the correlated and
conflicting data points mean that it is hard to estimate the
likelihood function in an objective way.  Our approach is to split the
wavelength regime in two at 10 \micron, since the values in the
optical/NIR and FIR are essentially unrelated.  In each of these
ranges, we take the likelihood to have a constant (top-hat)
distribution in log space between our minimum and maximum traces.  We
then convolve the two distributions together to describe the sum of
the flux in the two wavelength regions.

Of course, active galactic nuclei (AGN) also contribute to
the background energy.  This energy is emitted mainly in a UV/optical
peak, a roughly power-law X-ray tail, and an MIR/FIR peak from
dust-processed radiation.  
However, the AGN contribution to $\jbg$ appears to be much smaller
than that of star-forming galaxies.  For example,
\citet{phopkins06} present a detailed model of the AGN and supermassive
black hole populations that
fits (partly by construction) quasar luminosity function data over a
wide range of wavelengths and redshifts.
Their Figure~22b plots the growth of the average black hole mass
density with redshift, and we can compute the corresponding contribution
to the bolometric background radiation using their assumed
radiative efficiency of $\epsilon_r=0.1$
(essentially using the argument of \citealp{soltan82}).
We obtain $J_{AGN} = 1.9^{+1.5}_{-0.8}$ bgu, where the uncertainty 
is scaled from the uncertainty in their prediction of the $z=0$ black
hole mass density, 
$\rho_{BH} = 2.9^{+2.3}_{-1.9} \times 10^5 \msun \Mpc^{-3}$. 
Note that this argument does not assume that the \cite{phopkins06}
scenario is physically correct, just that it fits the observed
luminosity functions within their uncertainties and, therefore,
reproduces the observed total emission.
Furthermore, the model's predicted $\rho_{BH}$ agrees with observational
estimates (e.g., \citealt{aller02,marconi04}),
and there is no room for a large amount of additional AGN contribution
to $\jbg$ without overproducing these estimates.
A significant fraction of this bolometric AGN background
appears in X-ray wavelengths and therefore does not add to the
UV/Optical/IR background considered here.
As another check of $J_{AGN}$, the model of
\citet{silva04}, which is based on AGN counts in X-ray and IR bands
and has recently received further observational support
from \citet{treister06}, implies the AGN contribution in
all wavebands is $\tsim 1.4$ bgu, and that in the 
UV/Optical/IR background it is 1.1 bgu (Silva, private communication).
We conclude that the AGN contribution cannot be much larger than 4\%,
and is probably somewhat lower, and we ignore it henceforth.

The issue of whether or not the integrated galaxy counts match the total
EBL has implications for this paper beyond just a
change in the total bolometric background.  If flux from sources
besides normal galaxies are truly required to explain the extragalactic
light, and the source cannot be identified, then there is not 
necessarily a relationship between the total EBL and the energetic
output from stars!  However, the need for additional sources of energy
is not at all clear at present.  The systematic uncertainties,
particularly in the galaxy photometry and zodiacal light subtraction,
suggests that a curve splitting the difference between the raw counts
and measured total background (as in our ``mid'' model) represents the
most conservative assumption.  
Also, if the excess over total normal-galaxy counts should turn out to 
be real, a candidate for explaining the excess in the NIR,
where it appears most significant, exists in
the form of Population III stars.  According to models of these
sources, they would contribute almost all of their energy in the
wavelength range 1--5~\micron, leaving the background at other
wavelengths essentially unchanged.  This would imply that our lower limits
on the optical/NIR background (which ignore the NIR excess) and our
traces of the FIR background are still valid.

\subsection{Star Formation History}
\label{sec.obs.sfr}
The cosmic star formation history is related to both the EBL and the
stellar mass observable today.  Observational estimates of the SFH at
various redshifts are shown in Figure~\ref{fig.madau}.  This plot uses
a combination of IR (stars), UV continuum (filled symbols), and
emission-line surveys (open symbols).  We have tried to use the
deepest and broadest surveys within each type, so some of the
pioneering SFH surveys have been omitted.  The plot also shows the
sub-mm points from \citet[][crosses]{barger00}, including both their
actual source detections and their extrapolation to account for the
entire sub-mm background, as well as the estimate of the total SFR in the
damped Ly$\alpha$ population from \citet{wolfe03}.  We do not use the
sub-mm or damped Ly$\alpha$ points in the fit.  We omit estimates from
radio and X-ray surveys, because the energetic calibrators are even
more uncertain than those in the UV and FIR.

For the UV continuum measurements, we have combined results at
2800 \AA\ from \citet{connolly97} and \citet{wolf03},
and results at 1500--1700 \AA\ from
\citet{steidel99}, \citet{gabasch04}, \citet{giavalisco04}, 
\citet{schiminovich05}, \citet{sawicki05}, and \citet{bouwens06}.
For the IR measurements we include the local IRAS survey of \citet{yun01}
and the Spitzer 24 \micron\ survey of \citet{perez05}.
For emission-line measurements, we restrict ourselves to H$\alpha$
because metal lines like [O II] might be biased by evolution of the
metallicity and we use the values listed in
\citet{hopkins04} from the surveys of
\citet{gallego95}, \citet{tresse98}, \citet{yan99}, \citet{sullivan00},
\citet{hopkins00}, \citet{tresse02}, and \citet{perez03}.
We use only the H$\alpha$ estimate from \citet{sullivan00}
because the UV continuum magnitudes 
used there have been brought into question by more
recent results from GALEX \citep{schiminovich05}.

\madaufig

We convert the calibration factors used by the
various authors to the uniform values given in Table~\ref{table.sfrconv},
using a diet Salpeter IMF with $10^8 \yr$ bursts.  
We have converted all the star formation rate densities to our standard 
cosmology, since they scale with the Hubble constant, $H(z) = \dot{a}/a$.
This cosmology dependence factors out when integrating over time to
get $\rhoform$, although $\rhorem$ can shift by a percent or so owing
to the dependence of $f_{\it mass}$ on cosmic timescales.

The UV surveys, in general, do not probe deep enough in luminosity to
make the integrated star formation rate converge.  Most authors
calculate the total by fitting a Schechter form to the
luminosity function and extrapolating either to zero flux or to an arbitrarily
chosen lower limit.  For consistency, we adjust these reported values
to a uniform extrapolation down to $0.1 L_\ast$ where
$L_\ast$ is the Schechter turnover luminosity reported for each survey.
Extrapolating down to $0.01 L_\ast$ instead would cause the 
values to increase by 20--45\%.
In many cases ours
is the approach used by the original authors; in the
others the change in the extrapolation has a rather minor effect
compared to the scatter between surveys. 
In contrast to the UV, the H$\alpha$ surveys 
are deep enough that they require no extrapolation.

To make the UV continuum extinction corrections reasonably uniform, 
we use the correction formula of \citet{calzetti99}, assuming the 
mean value of $E(B-V) = 0.15$ given by \citet{steidel99}.  
This yields correction factors of
3.1 and 4.7 at 2800 and 1500 \AA, respectively.
The one exception is the highest-redshift point of \citet{bouwens06},
for which we have used the correction factor of 2.0 estimated by the
authors from the mean UV slope in the sample.  At lower redshifts,
some authors have used this UV slope correction method to obtain
higher extinction values than we have assumed (e.g., a factor 7 in both
\citealp{schiminovich05} and \citealp{giavalisco04}).  

For the H$\alpha$ measurements, we used the uniform luminosity-dependent
extinction calibration of \citet{hopkins04}.  The two FIR surveys are not
``transmission-corrected'' for the amount of UV light that leaks out
directly from the galaxies, but this is likely a significant fraction.
In fact, at low redshifts the UV and FIR luminosity estimates are roughly
equal \citep{martin05}.  We have crudely assumed a correction factor
of 2.0 for $z<1$ and 1.2 for $z\ge 1$.  
All of these extinction and transmission estimates are clearly highly
uncertain, and among the largest of the calibration errors in the
observed SFH.

\rhoderivfig

In Figure~\ref{fig.rhoderiv} we replot the SFH
such that one can easily integrate under the data points by eye to get the
total mass of stars formed:  $d{\rho_\star}/da$ versus scale
factor $a$.  This plot makes it straightforward
to see the relative importance of star formation at different
epochs in determining the stellar content today.
Clearly, most of the uncertainty in the total stars formed
comes from the redshift range $1 \lesssim z \lesssim 3$.  While there is 
still a surprisingly high dispersion in low-redshift ($z < 0.5$)
estimates of the star formation rate, the figure shows it is of little 
significance to the total stellar mass.

In these plots, it is unclear how well individual observations capture
the whole census of star formation at any redshift.  At high redshift,
several populations of sources have been suggested as major
contributors to the total star formation, including diffuse red
galaxies \citep{webb06}, damped Ly$\alpha$ absorbers \citep{wolfe03},
Ly$\alpha$ emitters \citep[e.g.,][]{tapken06}, and sub-mm galaxies
\citep{barger00,chapman05}.  For all of these categories, the extent
of overlap with the Lyman-break galaxies, for example, is still
unclear.  The mere presence of one population in a second population
may not suffice; for example, the large star formation rates found for
the bright sub-mm sources are essentially invisible to the SFR
estimates made in the UV \citep{chapman05}.  Hence, it is possible
that the SFR estimates obtained from these different populations
should be added together, rather than plotted together as different
measures of the same quantity.  This could raise the true SFR by
perhaps a factor of 2 above the individual points.

To compare this large spread of data points to other measures of star formation,
it is useful to obtain analytic fits to the star formation history.
Here we introduce a new parametric form: we set
\begin{equation}
\rhoform(z) = \rho_0 / (1 + p_1 a^{-p_2})^{p_3}
\end{equation}
where $a = 1/(1+z)$ is the scale factor.  Then
\begin{equation}
\rhodotform = \frac{p_1 \, p_2 \, p_3 \, \rho_0 \, a^{-p_2}}
{(1 + p_1 a^{-p_2})^{p_3+1}} H(z) \; .
\label{eqn.myform}
\end{equation}
This functional form has the advantages of easily relating 
$\rhodotform$ and $\rhoform$, and of 
making the dependence on cosmology explicit.  This parameterisation 
is a particularly natural choice for Figure~\ref{fig.rhoderiv}, since
$d\rhodotform/da = \rhodotform / \left[a H(z)\right]$ 
is a simple analytic function.  

We use a $\chi^2$ technique to fit the data.  The reported errors for
the different points vary widely, depending mostly on sample size; for
those with the smallest errors, the systematic errors in calibration
and extrapolation probably swamp the formal errors, which are derived
from Poisson statistics alone.  Rather than trying to revise the error
derivation of every literature source, we simply add an error of 0.1
dex in quadrature to the errors on the points before fitting the data,
which restores the best-fit $\chi^2$ to an acceptable value of 44 for
42 degrees of freedom.  The best-fit parameters are 
$\rho_0 = 9.0 \times 10^8 \Msun \Mpc^{-3}$, 
$p_1 = 0.075$, $p_2 = 3.7$, and $p_3 = 0.84$.

To test for a possible dependence of our results on the 
parametric form, we also use the form of \citet{cole01}:
\begin{equation}
\rhodotform = \frac{(a+bz) h}{1+(z/c)^d} \; .
\label{eqn.coleform}
\end{equation}
Here the dependence on the present-day Hubble constant is factored
out, although not that of the other cosmological parameters.  This
form has the same number of free parameters as the previous one and
has been widely used in the literature.  Fitting the data as before,
our best fit is
$a = 0.0103 \Msun \yr^{-1} \Mpc^{-3}$, 
$b = 0.088 \Msun \yr^{-1} \Mpc^{-3}$, $c = 2.4$, and $d = 2.8$ 
with $\chi^2=44.7$
for 42 degrees of freedom.
The curves produced by these two parameterisations 
are nearly indistinguishable from each other.

We want to probe not just the mean value but the full allowed range of
star formation histories.  To do this, we construct an ensemble of
parametric fits, using a Monte Carlo Markov chain and assuming uniform
priors on the logarithm of the parameters.  The ensemble evenly
samples the probability distribution of the fits.  We can then 
estimate the allowed ranges of $\rhodotform(z)$ at any $z$ from the
ensemble of fits.  (A simple grid search in parameter space was also
tried, and produced essentially identical results.)  The typical
formal $1\sigma$ error in $\rhodotform$ is only 10--15\% for $z < 5$,
as shown in Figure~\ref{fig.madau}.  The widths of these envelopes are
quite similar to those of \citet{hopkins06}, despite the different
data sample, sampling procedure, and parametric form.

However, it is likely that this procedure underestimates the true
errors in the determination of $\rhodotform$.  For example, the best
fit in \citet{hopkins06} appears to lie well above our error envelope for
$2<z<4$, owing to a different selection of data points, different
assumptions about extinction, and a larger extrapolation of the
luminosity function.  While the large number of data points with small
Poisson errors result in a small formal error in our fit, this masks
the larger uncertainties in calibrating $\rhodotform$, which applies
equally to all the points.  Even with a fixed IMF, there are
substantial uncertainties in the mean opacity at a given redshift and
in the calibration of luminosities owing to the uncertain star
formation histories of individual galaxies (see \S 2).  The systematic
errors at different redshifts are undoubtedly correlated, although
there is speculation that the mean opacities at $z \sim 0$, $z \sim
2$, and $z > 4$ are substantially different from one another.

Therefore, we produce an alternate grid of fits in which we randomly and
smoothly adjust the $\rhodotform$ data values as a function of
redshift.  We first define three random variables $A(0)$, $A(1.5)$,
and $A(5)$, and set each to have normal distributions of dispersion
$\sigma_A=0.5$.
We then multiply the data values and associated errors by
$e^{A(z)}$, where $A(z)$ is the quadratic function in $z$ through the three
points $A(0)$, $A(1.5)$, and $A(5)$, and repeat the previous
Monte Carlo sampling procedure.  Not surprisingly, this procedure
results in a much broader range of acceptable parameters, and a
correspondingly wider envelope of $\rhodotform$ at any given redshift,
as shown in Figure~\ref{fig.madau}.  Given the large uncertainties
discussed above, $\sigma_A = 0.5$ is probably reasonable.
The 1-$\sigma$ dispersion in the SFH is then about 50\% or 0.2 dex for $z < 4$.

\subsection{Stellar mass and NIR luminosity density}
\label{sec.obs.mass}

The third observable we need to consider is the stellar mass density.
We are primarily concerned with the value of this quantity today but
we shall also consider the evolution of stellar mass with cosmic time.
NIR light is a useful proxy for stellar mass, since plausible values
of $M/L$ vary only by a factor of a few and this variation can be
accounted for using simple colour-based estimates \citep{bell01}.
Like the star formation rate, the stellar mass and luminosity densities 
determined from observations are proportional to $H(z)$, and we have
translated the measurements below to our assumed cosmology.

We shall use five studies of NIR light and stellar mass at low
redshifts.  All the studies in the $J$ and $K$ bands use 2MASS
magnitudes, combined with various redshift surveys.  The results
of these local surveys are given in Table~\ref{table.localmass}.

\citet{cole01} use the 2dF redshift survey in the southern sky,
combined with the second incremental release of 2MASS, to obtain
$J$, $K$, and stellar mass densities.  They use Kron
aperture magnitudes, calibrating the preliminary 2MASS catalogue
against deeper $K$-band data and correcting their estimate of
$\rhorem$ for missing light outside the Kron radius.  However, they do not
similarly correct their reported estimate of $\rho_K$, so we have
scaled up their Kron magnitude based estimate by 0.11 magnitudes to
get the total light.  To determine the stellar mass density they estimate the
mass-to-light ratio galaxy by
galaxy using the observed optical-to-near-infrared colours to
constrain the effects of the star formation history and metallicity in
each galaxy.  They check for the effect of missing low-surface-brightness
galaxies in the relatively shallow 2MASS sample and conclude that that any
such effect should be small.  

\citet{bell03} use the Sloan Digital Sky Survey (SDSS) Early Data
Release in the northern sky, combined with the final 2MASS survey,
to obtain $K$-band and stellar mass densities.  They also use Kron
magnitudes but scale up the luminosities of the early-type
galaxies by 0.1 mag to account for light outside the Kron aperture.
They estimate their $M/L$ ratios by means of optical galaxy colours; they neglect
the effect of extinction entirely since it affects the mass estimate
both through the flux and the colour, and these should nearly cancel out
\citep{bell01}.  They go on to suggest that the 2MASS catalogue may miss some
low-surface-brightness galaxies and estimate the magnitude of this effect in
three ways.  First, estimating the $K$-band light density using the $r$-band
luminosity density, which should be complete at these magnitudes, and the
global $r-K$ colour for their sample, they find $\rho_K$ increases by 12\%.
Second, using their $g$-band-selected galaxy sample and synthesising the $K$
magnitudes of galaxies missing from the $K$-band sample from their
optical magnitudes, they find $\rho_K$ increases by as much as 30\%.
This $g$-band estimate might be biased since
when they obtain the stellar mass density using the complete $g$-band optical
sample it is only 4\% higher than their estimate using their uncorrected
$K$-band magnitude limited sample.  Hence, in this paper we use their 
direct estimates of mass and light from the $K$-band sample.

\citet{panter04} analyse galaxy spectra in the first data release of
the SDSS at 20 \AA\ resolution.  They obtain total fluxes by
normalising the spectroscopic fibre magnitudes to photometric
$R$-band Petrosian magnitudes.  (We note that Petrosian magnitudes may
underestimate the flux from early-type galaxies by about 20\%, although
they are probably very accurate for disk galaxies \citep{strauss02,graham05}.
They then fit galaxy models to compressed versions of the spectra
to obtain the local stellar mass density.  

\citet{eke05} use an updated version of the 2dF redshift survey
together with the updated 2MASS catalogue to produce a larger sample
than \citet{cole01}, obtaining $J$-band, $K$-band, and stellar mass densities.
Eke \ea calculate galactic stellar masses using star formation histories that
include bursts. When
tested against mock galaxy catalogues, this reduces the average stellar masses
by 20--30\%.  Whether or not this results in a smaller bias, however, depends on
the validity of the star formation histories in the mock catalogues.
Eke \ea show that the mass function obtained from their
optically-selected sample differs systematically from that obtained
using the $J$-band selected sample, which they cite as possible support for the
bias against low-surface-brightness galaxies found by
\citet{bell03}.  However, the difference starts starts to occur
where the computed biases in the optically-based and NIR-based mass
estimates from the mock catalogues start to diverge, making its significance
unclear.

Finally, \citet{jones06} combine 2MASS with a preliminary catalogue
from the wide-angle 6dF redshift survey and determine $J$-band and $K$-band
luminosity densities.  They use ``total'' (extrapolated) 2MASS
magnitudes to obtain the galaxy luminosities.  Since they do not explicitly
state a error estimate, we read it off their Figure 15.
They also argue that their sample is robust to surface brightness effects
on the total luminosity density.

\localmasstable

For our purposes we must compare the different estimates using the same
assumptions.  First, we compute the luminosity densities for each estimate
using the same value for the sun's
absolute magnitude in each band, obtaining the densities in
Table~\ref{table.localmass}.  Since the luminosity densities were computed
in a similar way by each group, we first average the luminosity
density estimates after weighting by the inverse of its variance, and take the
harmonic sum of the variances to get the statistical error.  We then
add an overall systematic error of 15\% (cf.\ \citealp{bell03}), which
dominates the error in all cases.

To compute the stellar mass densities,
we first convert all the mass estimates to a
diet Salpeter IMF, assuming 0.7 times as much mass as for a Salpeter
IMF and 1.74 times as much mass as for a Kennicutt IMF.  We then take
a straight average of the four estimates, since for each data point 
the systematic error outweighs the statistical error
and we do not know which approach is best.  
We then assign an overall systematic error of $20\%$ (cf.\
\citealp{bell03}), which may be reasonable 
as the varied approaches here yield a dispersion of only 10\%.
Our final estimate of $(3.50 \pm 0.70) \times 10^8 \msun \Mpc^{-3}$ 
for a diet Salpeter IMF
implies $\Omega_\star h = (1.8 \pm 0.4) \times 10^{-3}$ and 
$F_\star = \Omega_\star/\Omega_b= 0.063 \pm 0.016$.  
This estimate is lower by
more than a factor of two than the earlier best estimate of
\citet{fukugita98} using $B$-band data, once we account for the
different IMF.

\starmasszfig

Recently, attempts have been made to estimate the stellar density at
higher redshift as well, based on faint galaxy catalogues with
multi-band photometry.  This is even more difficult than estimating
the high-redshift star formation rate density, both because surface
brightness dimming interferes more with regions of less intense
emission and because their estimates must be based on the rest-frame
optical light rather than rest $K$-band, which increases the uncertainty in
the mass.  In Figure~\ref{fig.starmassz} we show a number of
observational estimates for $\rho_\star(z)$.  Most of these points
were originally derived for a Salpeter IMF.  We rescale them to a diet
Salpeter IMF using the ratio of population masses calculated from our
best-fit star formation history; we find this to be a factor of 0.70--0.74.
The different studies use a variety of assumptions about stellar populations,
extinction, and methods for calculating the mean $M/L$ ratios.  
In most cases we plot the error bars given by the authors, which are usually
purely from Poisson statistics.  However, for
\citet{dickinson03} we plot error bars based on their dominant uncertainty, 
modelling the stellar population.  The largest estimates of
$\rho_\star$ at high $z$ are from \citet{gwyn05} but it is not clear
whether or not these higher estimates arise from their use of deeper UDF
data.

The systematic errors in the measurement of the total stellar mass and
light deserve some further discussion.  Extrapolation beyond the faint
or bright end of the observed luminosity function is not a significant
factor in the surveys of the local universe discussed above \citep{cole01,panter04}.
These surveys make corrections for flux outside the apertures
and these apertures are quite wide to begin with, e.g.
$\tsim 20$--$40 \kpc$ in radius for a massive disk galaxy, but
these corrections do not take into account the power-law halos that
commonly surround galaxies \citep{mouhcine05,raja05}.  However,
using the mass fraction and density profile suggested for M31 by
\citet{raja05}, it appears the extra correction would only amount
to 2\%.

Free stars not clearly bound to any single galaxy could be another
source of error.  Observations of red stars \citep{durrell02} and
planetary nebulae \citep{okamura02} suggest free stars make up
10--20\% of all the stars in the Virgo cluster.  However, as clusters
contain only about 2\% of all the stars in the universe \citep{eke05}, it by
itself is not a significant correction.  In simulations of galaxy formation
presented in a related paper (Fardal \ea 2006, in prep), free stars
are found preferentially in and around galaxy clusters owing to
the intense tidal interactions present there, so once again
the global fraction of free
stars should be well below that estimated for Virgo and be negligible
for our purposes here.

Systematic effects from low-surface-brightness galaxies unrepresented
in the 2MASS catalogue are potentially a larger problem; the various
estimates of their significance in the discussion above range from 0
to 30\%.  We will assume that this is not a significant bias, but the
reader should keep in mind that the true luminosity and mass densities
could potentially be raised by as much as $\tsim 30\%$ by this effect.

Finally, the calculation of $M/L$ ratios of individual galaxies, something
which is required in all these papers,
is uncertain owing to many factors including uncertainties in: the
metallicity, the star formation history, the extinction, the stellar tracks,
and the population synthesis methods themselves.  The
derived distribution of the metallicity may also have a substantial
effect, particularly on the $K$-band light from RGB and AGB stars.
The latter stars are quite difficult to simulate, and regardless of
the metallicity they are a major source of error in the stellar
population calculation.  To indicate the dispersion, the different
methods of \citet{cole01}, \citet{bell03}, and \citet{eke05} give
$M/L_K$ estimates of 0.88, 0.92, and 0.62 respectively, and our
combined sample gives a net result of 0.83.  This range of estimates
is in accord with the $\tsim 25\%$ systematic errors in $M/L$ (for a
fixed IMF) estimated by \citet{bell03}.  Observationally testing the
distributions of star formation, extinction, and metallicity derived
in local surveys will help control the systematic errors involved in
calculating the stellar mass.

\section{COMPARISON OF STAR FORMATION MEASURES}
\label{sec.compare}

We now turn to the central question of this paper: are the direct
estimates of the cosmic star formation history, the measured
bolometric background intensity, and the local density of $K$-band light
mutually compatible, given the assumption of a universal IMF with a
Salpeter-like shape above $0.9 \Msun$?  The curves in
Figure~\ref{fig.starmassz} show the stellar mass density as a function
of redshift computed from the star formation history traces shown by
the corresponding curves in Figures~\ref{fig.madau} and
\ref{fig.rhoderiv}.  The lowest curve, which reduces the standard
normalisation of the SFR estimates by about a factor of two (see \S
3.2), roughly follows the median of the data points, while the other
curves are systematically higher than the data.  We have included the
slight variation with redshift of the recycled mass fraction in
this calculation.  Although many factors could contribute to systematic
errors, it is of particular note
that a substantial amount of stellar mass could be
missing from the high-redshift observational surveys owing to surface
brightness effects.  
The tension between the SFR and stellar mass density estimates
provides circumstantial evidence for such effects.  If lower surface
brightness galaxies are indeed missed in the high redshift counts,
it would help resolve the discrepancy between the absolute background
light estimates and the integrated count estimates, suggesting
that the higher, absolute estimates are more accurate.

For the remainder of this section, we will focus on $z=0$, since the
observational estimates of the stellar mass density are more secure in
the local universe and, even more important, we can bring in the
bolometric background (which is only measured at $z=0$) as an
additional observational constraint.  We will mainly focus on the
local $K$-band light density $\rho_K$ rather than the stellar mass
density $\rhorem$, since it is the directly observable quantity.  We
integrate each SFH fit to obtain the predicted final stellar mass
$\rhorem$ and $K$-band luminosity density $\rho_K$ as in
Equation~\ref{eqn.rhostar}, and the EBL, $\jbg$, as in
Equation~\ref{eqn.backgd}.  (The integrations cover the range $z \leq
15$ in each case.)  We include the extinction correction
$e^{-\tau_K}$ discussed in \S\ref{sec.calib} to make $\rho_K$ an
observed luminosity density.

According to our direct SFR fits (assuming no calibration error), the
mean $\rhorem = (5.3 \pm 0.3) \times 10^8 \msun \Mpc^{-3}$, or
$F_\star = 0.095 \pm 0.005$.  The corresponding predicted 
extinction-corrected
$\rho_K = (6.9 \pm 0.4) \times 10^8 \,L_{\sun K} \Mpc^{-3}$. (We
obtain the same results whether we use our new parametric form in
Equation~\ref{eqn.myform} or that of \citet{cole01} in
Equation~\ref{eqn.coleform}.)  
Including an extra calibration error with $\sigma_A=0.5$, we obtain
$\rho_K = (6.9 \pm 2.2) \times 10^8 \,L_{\sun K} \Mpc^{-3}$.
Our adopted value for the observed density is
$(4.2 \pm 0.6) \times 10^8 \,L_{\sun K} \Mpc^{-3}$.  Thus the
integrated SFH should produce more mass and light than that directly
observed, by a factor of 1.5 if we use the quoted mass estimates, or
by a factor 1.6 if we use the $K$-band luminosity density.  The
probability of agreement is $< 2\%$ without random calibration
errors, but rises to 10\% and 32\% 
for random calibration error of $\sigma_A = 0.2$ and 0.5 respectively.  
Thus, while the offset could indicate that local surveys miss significant 
amounts of stars, or that some other assumption (such as our IMF) may be wrong,
calibration uncertainty in the SFH seems a sufficient explanation.
For example, the curve could be made to match the plots by
changing the estimates of the UV extinction.
Our best SFR fits imply $M/L_K = 0.71 e^{\tau_K} = 0.77$,
which is within the range of values found by \citet{cole01},
\citet{bell03}, and \citet{eke05}.

The appearance of a discrepancy is found by other authors as well,
although a number of authors apparently simply scale the integrated
SFH to match the $z=0$ stellar mass, without mentioning the scaling
factor!  Our integrations of $\rhorem$ are in reasonable
agreement with the results of \citet{hopkins06}, who found about a
10\% uncertainty in the integrated stellar mass from their star
formation fits and a value exceeding that expected from direct
observations by nearly a factor of 2 (for comparisons using the same
IMF).  The discrepancy implied by their final mass is even larger than
in our case, because their much larger SFR in the range $2<z<4$
contributes a significant amount of mass.

However, we can also include the EBL in the comparisons.  
Our adopted allowed range of $\jbg$ is 50 to 129 bgu, with a most
likely value of 77 bgu.  Our derived value from the SFH fits is
$(63 \pm 4)$ bgu without including random calibration error, and
$(62 \pm 20)$ bgu with $\sigma_A=0.5$.

\kbgcontours

After allowing for fairly large systematic uncertainties, it appears 
that the estimated cosmic star formation history 
can be made consistent, individually, with
measurements of the local $K$-band light density and the bolometric
background intensity.  One might, therefore, be tempted to conclude that
these three probes of the cosmic star formation are mutually
consistent.  Figure~\ref{fig.kcontours} shows that this is not the
case.  In the upper left panel, we show acceptable fits in the space
of $\rho_K$ vs.\ $\jbg$, assuming no systematic errors in the
calibration of the SFRs.  The vertical lines represent the mean and
$2\sigma$ errors in the local $K$-band luminosity density from the
combined observations, as discussed in \S~\ref{sec.obs.mass}.  
The horizontal lines represent the three
values of the background derived in \S~\ref{sec.obs.backgd}.  
The rounded, vertically extended contours
represent joint confidence intervals for these two parameters.  We
construct these by multiplying the probability distributions of the
two parameters together, and finding the contours that enclose 65\%,
95\%, and 99.8\% of the probability.  The ``squashed'' appearance of the
contours comes from the non-Gaussian probability distribution of the
background light (see \S \ref{sec.obs.backgd} for an explanation of
how we estimate its distribution).

The diagonal contours in panel (a) depict the 65\%, 95\%, and
99\% 2-dimensional confidence regions derived from the ensemble of star
formation histories.  These contours form a narrow band.  If we
consider only one dimension, our computed values of $\jbg$ are in
reasonable agreement with observations, while the stellar mass is
rather high as mentioned before.  The main point, however, is that
the two sets of contours barely overlap; one needs to go out to the
$3\sigma$ contours to find agreement.
We obtain essentially identical results if we use the \citet{cole01}
parametric form for the star formation history instead of the form in
Equation~\ref{eqn.myform}.  

The remaining two panels show the results for samples with increasing
levels of assumed systematic error in the overall calibration of the
SFRs, i.e. a random dispersion in the calibration of $\rhodotform$.  This
greatly expands the allowed joint confidence region of $K$-band light and
EBL so that, as we just discussed, the former is no longer
inconsistent with the observations.  However, the confidence region
still forms an extremely narrow band, which overlaps the observational
contours at only the 2 or $3\sigma$ level.  Even with this increased
freedom, {\em one can find SFR histories that agree well with either
  the observed $K$ light or the observed bolometric background, but
  not with both quantities simultaneously.}

\mbgcontours

We consider other proxies for the stellar mass as well.  Plotting
$J_{\it EBL}$ versus $\rho_J$ (instead of $\rho_K$) in
Figure~\ref{fig.masscontours}, we see that the $2\sigma$
contours still do not overlap.  If instead we use the SDSS $z$-band density,
$\rho_z$, the contours still do not significantly overlap.
The slightly better agreement in $z$-band might suggest that
there may indeed be a small bias in the 2MASS surveys owing to missing
low surface brightness galaxies.  However, our extinction estimates become
increasingly suspect as we go to these shorter wavelengths.
Finally, we plot the contours of stellar mass itself.  With the larger
observational uncertainties in this quantity, the disagreement is now
at less than the $2\sigma$ level, though the impression remains that
either the observed background is too high or the observed mass too low.  We
regard the $K$-band plots as more fundamental, since they plot a
directly observed quantity and one that is tied as closely as possible 
to the dominant stellar mass range.

In essence, we find that the bolometric background is too bright
relative to the present day $K$-band light density, given a
Salpeter-like IMF and the shape of the SFH implied by our fits to
the observations.  Changes to the normalisation of the SFH move
predictions along the diagonal in Figures~\ref{fig.kcontours}
and~\ref{fig.masscontours} and, therefore, do nothing to resolve
this discrepancy.  Our $\sigma_A=0.5$ models allow substantial
variations in the SFH shape as well as normalisation, but
these have limited power to solve the problem.
The ratio between the background production
function $S_{\it EBL}(z)$ and the $K$-band luminosity $\rho_K(z)$ has a
maximum at $z \approx 1$ and falls off slowly at higher and lower redshifts.
Hence, by the mean value theorem, the ratio between the EBL and the emitted
$K$-band luminosity density cannot be larger than
\begin{eqnarray*}
\max\left( \frac{\jbg(z)}{\rho_K(z)}\right)
&=& \frac{c}{4 \pi \, e^{-\tau_K}}  
  \max\left( \frac{S_{\it EBL}(z)}{l_K(z)} \right) \\
&=& 1.21 \times 10^{-7} \mbox{bgu} 
                        [L_{\sun K} \,\Mpc^{-3}]^{-1} \; .
\end{eqnarray*}
for a Salpeter or diet Salpeter IMF, including the extinction correction.  
We plot the maximum ratio as the diagonal dotted line in
Figure~\ref{fig.kcontours}.  Because much of the star formation
in our calculations takes place at $z \sim 1$, 
the predicted ratio for our best-fit SFH, 
$9.0 \times 10^{-8} \mbox{bgu} [L_{\sun K} \,\Mpc^{-3}]^{-1}$,
is already 75\% of this upper limit,
so even a $\delta$-function SFH at $z=1$ can produce only
a modest increase.
The ratio of our middle trace EBL to our best estimate $\rho_K$ is
$1.8 \times 10^{-7} \mbox{bgu} [L_{\sun K} \,\Mpc^{-3}]^{-1}$.
Getting down to 
$1.0 \times 10^{-7} \mbox{bgu} [L_{\sun K} \,\Mpc^{-3}]^{-1}$
requires taking our minimal estimate of 50 bgu for $\jbg$ 
(the lower trace in Figure~\ref{fig.obsbackgd}) and increasing
our averaged value of $\rho_K = 4.2 \times 10^{8} L_{\sun K} \,\Mpc^{-3}$
by 15\%.

With our best SFH fit, 50\% of the background is produced in stars
that form at $z < 1.0$, and the central 80\% comes from the stars that form
in the redshift range $0.27 < z < 2.6$.  This can be compared with the
present-day stellar mass, for which 50\% comes from $z < 1.2$, and the central
80\% comes from the redshift range $0.33 < z < 3.2$.  For the $K$-band
luminosity density, 50\% comes from stars formed at
$z < 0.9$, and the central 80\% comes from
the redshift range $0.1 < z < 2.7$.  Therefore, the EBL principally comes
from stars formed at the same redshifts as those that dominate the
current mass or $K$-band light, which explains why the correlation is so
tight.  The background energy is {\it emitted} at still lower
redshifts.  Again, using our best estimate of the SFH, 50\% comes from
$z < 0.65$, and the central 80\% comes from the redshift range $0.11 < z < 1.9$.

Even if we consider changes in the star formation history that are
more radical than those allowed by our random-calibration formalism,
it is still difficult to change the ratio of background light to
stellar mass significantly.
The ``fossil record'' SFH, estimated from the
spectra of local galaxies by \citet{heavens04}, peaks at a
significantly lower redshift than our typical fits, which is more
favourable to the creation of EBL.
However, the ratio of $K$-band light to stellar mass, using this star
formation history, would rise by only a few percent and the ratio of
EBL to $K$-band luminosity would actually drop by a few percent,
since stars at lower redshifts are even more capable of producing $K$-band
light.  Two independent calculations of the EBL versus stellar mass from
the literature, which we discuss below, are also shown in
Figure~\ref{fig.masscontours}. These values bolster our conclusion
that the ratio is insensitive to reasonable changes to the SFH.
Since all allowable star formation histories
imply that most stars are old (Figure~\ref{fig.rhoderiv}),
even significant shifts in the distribution of their formation
redshifts only results in minor changes to the majority of the stellar
ages and to the resulting stellar energy output.

The values of $\rho_\star$ and $\rho_K$ calculated from the
observations are proportional to the value of the Hubble constant,
$H_0$.  However, the observed EBL and $\rhoform$, integrated from the
SFH, are both independent of the Hubble constant, as is
$F_\star$ also computed from the integrated $\rhoform$ owing to the
invariance of $\rho_{\it baryon}$ derived from microwave background
measurements.  So an uncertainty of $7 \kms \Mpc^{-1}$ in our
canonical value of $H_0 = 70 \kms \Mpc^{-1}$ will add a 10\% uncertainty
when comparing $\rho_\star$ to these other quantities.  (The changes
from offsets in the lookback-time arguments in Equations 1--3 are
smaller still and can be neglected.)

In Figure~\ref{fig.kcontours}, we have not taken into account the
systematic error from uncertainties in the Hubble constant or other systematics
such as uncertainties in the stellar population tracks and synthesis, the
stellar metallicity, or the mean NIR extinction.  Taken together,
these systematics may combine to a net 25\% error, but we do not
place much faith in this estimate.  With
systematics in mind, an observational solution to the background
problem seems reasonable, perhaps giving values close to 
$\rho_K = 5.0 \times 10^8 \Msun \Mpc^{-3}$ and $\jbg = 55$ bgu.  
We emphasise that for the background light to be so near its lower
limit, a large set of absolute background detections would need to be
wrong, and if future observations substantiate the higher detections in
{\it any} region of the spectrum, 
an ``observational'' solution will be essentially ruled out.

Our conclusion of a potential conflict runs contrary to that of
\citet{madau00} and \citet{madau01}, who find very good agreement
between the background radiation and the stellar density for their
assumed star formation histories.
\footnote{The integrated background from the
  ``realistic'' star formation history in \citet{madau00} appears to be
  incorrect, but this is corrected in the later conference proceedings
  \citep{madau01} along with slight changes to the assumed star
  formation history.  The ratio of background radiation to 
  stellar mass density for the cases considered in the latter paper
  is then quite insensitive to the star formation history, in
  agreement with our results above.}  
These authors use a stellar density of $\rho_\star = 6.8 \times 10^8
h \msun \Mpc^{-3}$ based on B-band surveys, and find a best EBL value of
$\jbg = 55$ bgu \citep{madau00} or 60 bgu \citep{madau01}.  We plot this result
in Figure~\ref{fig.masscontours}, after translating to our IMF and
including mass recycling, and it is in good agreement with our {\em
ratio} of $\jbg$ to $\rho_\star$.  Since that time, the estimates of
$\rho_\star h^{-1}$, using larger and more accurate NIR-based
galaxy surveys, have dropped by nearly a factor two, a change only
partially mitigated by the change in the value of the Hubble constant from
their assumed value of 50 $\kms \Mpc^{-1}$ to our assumed 
value of 70 $\kms \Mpc^{-1}$.  Also, additional observations of the
background light have raised the most likely value of $\jbg$, making
the previous ``best values'' closer to our lower limit.  These
two changes explain their discrepancy with our conclusions.
 
\citet{nagamine00} computed the star formation history using an LCDM
hydrodynamic simulation, and find from this history that
$\jbg = 81$ bgu and that the stellar density, using a Salpeter IMF,
is $\Omega_\star = 0.012$.  Translating this to our diet Salpeter IMF and
including mass recycling implies
$\rhostar = 8.0 \times 10^8 \msun \Mpc^{-3}$, over twice as large as
what is acceptable today.  We also show this result in
Figure~\ref{fig.masscontours}.  
Nagamine et al.'s {\em ratio} of $\jbg$ to
$\rhostar$ is only 15\% lower than ours, and this slight difference
probably owes to their high level of star formation at late times,
which exceeds observational constraints.  
In a more recent paper (which appeared in preprint form as we
were finalising this manuscript for submission), 
\citet{nagamine06} present empirical and numerical models of
the cosmic SFH which, they argue, are consistent with both
the extragalactic background light and the local bolometric
luminosity density.  However, they adopt a background value
of $\jbg \approx 45\,$bgu, which is even lower than the
value of 50 bgu implied by our lowest trace through the
data in Figure~\ref{fig.obsbackgd}; in essence, they assume
that the integrated count estimates of the background are
correct and that all of the absolute measurements are incorrect.
We also regard the local $K$-band light density as a more 
robust ``fossil'' constraint on the SFH than the local bolometric
light density, because the latter is more difficult to estimate
observationally and is more sensitive to contributions from
very recent star formation.  We concur with \citet{nagamine06}
that the directly estimated SFH, extragalactic background light,
and local light density can be reconciled if one takes a 
low value for the second and a high value for the third,
but this solution requires pushing the systematic uncertainties
of the observations to their limits.

\section{SOME POSSIBLE SOLUTIONS TO THE BACKGROUND PROBLEM}

\label{sec.discuss}

With our assumed diet Salpeter IMF, the local stellar mass density
(or light density) appears to be inconsistent with the directly
estimated cosmic star formation history and the extragalactic
background light.  The local light suggests a stellar baryon fraction
$F_\star \approx 0.06$, while the
integrated star formation history suggests a higher value of
$F_\star \approx 0.09$. The EBL suggests values that
are similar or even larger, $F_\star \gtrsim 0.08-0.20$, and the fairly
hard lower limit on the observed EBL makes it difficult to accommodate without
making the stellar mass density too high.  
The calibration of the SFH is uncertain enough that it can be made
consistent with either of the other two constraints individually,
but not simultaneously.
Systematic uncertainties in the absolute background measurements
and the completeness of the local galaxy census leave room
for an ``observational'' solution to this conflict.
However, if future measurements accurately constrain
the stellar mass density and background light to the most probable values we
have obtained here, there will be a substantial discrepancy.  

Let us, therefore, consider the alternative solution of extra energy 
in the background beyond that provided by a Salpeter IMF.
Solving the background problem using an additional
source of energy requires an extra 10--40 bgu, or
0.4--$1.7 \times 10^{-14} \erg \cm^{-3}$, which is comparable to the energy
density of stellar light from all the detected galaxies, even discounting the
effect of redshift.  This amounts to $\tsim 10^{-6}$ of the closure
energy density, or about 12--50 keV per baryon.  This is a steep
requirement for any possible energy source (see \citealp{fugukita04}
for a useful list of candidates).  For exotic sources, the requirement
that the excess appear in the optical/NIR or FIR portions of the
spectrum is an additional hurdle.

The energy requirement alone immediately rules out most possibilities,
including supernova-driven shocks and gravitationally powered cooling
radiation.  Decaying cosmic neutrinos or decaying dark matter would
have sufficient energy, but producing the required decay rate and
appropriate photon energies would require substantial fine-tuning.  

AGN are a known contributor to the background at the $\tsim 2$ bgu
level, as discussed in \S~\ref{sec.obs.backgd}.  A large (e.g.,
factor of 5-10) increase in this contribution is possible only
if the local census of fossil black holes is seriously incomplete,
perhaps because the great majority of supermassive black holes
have been ejected from galaxy centres in mergers.
A number of authors have suggested Pop III stars or their
black hole remnants as sources of the high background measured
in the $1-4\mu$ region, but \citet{madau05} show that the 
energetic and chemical constraints on such a population are
difficult to satisfy; furthermore, they have been suggested
to explain a set of measurements that we are already not
including in our fiducial estimate of $\jbg$.

Thus, non-stellar solutions to the background problem do not appear
promising.  On the other hand, we have not yet considered 
modifications of the IMF.  An IMF that is biased to
intermediate- or high-mass stars, compared to the diet Salpeter IMF we
have used up until now, could have a significant effect on the stellar
emission.  These stars would mostly burn out by the present day, thus
contributing to the observed SFH and EBL but not to the $K$-band light density.
As noted in the introduction, improving agreement among these
observations requires changing the IMF {\it above} the turnoff mass
$0.9M_\odot$ of old stellar populations; changes below this mass
renormalise all three quantities by the same 
factor.\footnote{\citet{nagamine06} conclude that IMF changes have
little impact on the predicted EBL, but that is because they only compare
two cases (Salpeter and Chabrier) that are nearly identical 
above $1 \Msun$.}

\imfplot

In Figure~\ref{fig.imfplot}a
we show several IMFs that have been proposed in the
literature \citep{salpeter55,millerscalo79,kennicutt83,kroupa93,
  kroupa01,chabrier03,baldry03}, all normalised to the same value at
$1 \msun$.  We also show
the approximate allowed region of the IMF as determined by \citet{kroupa02}.
This allowed region was established from studies using several different
techniques covering different mass ranges; hence the normalisations
above and below $1 \msun$ are poorly constrained relative to one another.
The slopes from 1 to $15 \msun$ in the plot are 1.7 for Kroupa
(1993), 1.6 for Miller-Scalo (1979), 1.5 for Kennicutt (1983), 1.35
for Salpeter (1955), 1.3 for Kroupa (2001, mean Galactic-field form in
Equation 2) and Chabrier (2003), and 1.2 for \citet{baldry03} IMFs.

Now let us repeat the exercise leading to Figure~\ref{fig.kcontours} but
using the different IMFs.  We normalise the star formation histories by the
amount of UV light produced in a $10^8 \yr$ burst.  (Using the
bolometric luminosity at the same burst length would give a similar scaling.)
Figure~\ref{fig.imfplot}b shows the resulting $K$-band-$\jbg$ contours.
There is a single overall trend in this plot.  The stars
that dominate the $K$-band light today ($\rho_K$), the EBL, and the UV light are
of progressively higher mass, so as the IMF slope above $1 \msun$ is made
shallower, the contours shift to lower values of $\rho_K$ and $\jbg$, while at
the same time the ratio of $\jbg / \rho_K$ increases.  (Changes to the
IMF below $1 \msun$ have little effect on the $K$-band or bolometric
light.)  Because $\rho_K$ is most likely better known than the
calibration of the SFH, making the IMF shallower above $1 \msun$
actually {\it raises} the most plausible background value, despite the
resulting decline in the mean $\jbg$ of the contours.

We consider the Salpeter IMF to be the point at which the agreement is
marginally acceptable, and rule out IMF slopes steeper than the
Salpeter value of $-1.35$.  This excludes a number of IMFs commonly
used in the literature.  The \citet{baldry03} IMF is the most
acceptable; these authors estimate a high-mass slope of $-1.15
\pm 0.2$, which independently rules out the steepest IMFs. Alone among
the IMFs in Figure~\ref{fig.imfplot}b, the Baldry-Glazebrook IMF
was derived from the global SFH, in this case by comparing the
spectrum of the local luminosity density to observationally plausible
shapes of the SFH.  It is interesting that it agrees best with
our independent test, perhaps indicating that the
estimates from Galactic data are somehow biased or that the Galactic
IMF differs from the Universal average.

However, {\em none} of the standard IMFs from the literature give more
than marginal agreement with our joint contours in $\rho_K$ and
$\jbg$.  This will pose a problem if future observations constrain the
background to a high value.  We now consider several examples to see
what is required to get significantly higher levels of background light.

We first consider the case of universal IMFs. For our purposes this does not
necessarily imply a completely invariant IMF from one galaxy
to another, but does require that there be no systematic change
with redshift.  It is useful to distinguish between
a top-heavy IMF, rich in high-mass ($M > 8 \msun$) stars relative to
Sun-like stars, and a middle-heavy or ``paunchy'' IMF, rich in
intermediate-mass ($1 \msun < M < 8 \msun$) stars.  As discussed above, the
particular range that would be most efficient in boosting $\jbg$
relative to the UV/FIR or $K$-band light is 1.5--$4 \msun$.

In Figure~\ref{fig.imfplot}c and Table~\ref{table.imf} we introduce three
additional IMFs.  The first is an ``extreme top-heavy'' IMF
(the slope here is $0.95$, so that the mass converges only at the
upper end of the mass function).  As an example of a
``paunchy'' IMF, we place a break at $4 \msun$, and for a more
extreme or ``obese'' example, we place a discontinuity at $1.5 \msun$.
For comparison, we also include a diet Salpeter IMF,
and we use the Baldry-Glazebrook IMF as an example of a modestly top-heavy
IMF.

We plot results in the $\rho_K$-$\jbg$ plane for these IMFs in 
Figure~\ref{fig.imfplot}d. Here, we only
represent the results in a schematic way to avoid overlapping
contours.  The top-heavy examples continue the general trend down and
to the left in the plot.  The paunchy examples, by contrast, go only
to the left since they have enhanced bolometric emission relative to the
$K$-band or UV light.

Our paunchy and extreme top-heavy IMFs reside within the hatched allowed
regions of \citet{kroupa01} and \citet{kroupa02}.  For 0.5--$1.0
\msun$, the paunchy IMF is in marginal conflict with Kroupa's limits
on the slope, but this is based on only a few samples with large
error bars and potentially large systematics.  The Baldry-Glazebrook
and extreme top-heavy IMFs are in good agreement with all of these limits.
The obese IMF, however, is too ``paunchy'' to be consistent with 
Milky Way observations.

The need to join together estimates in different mass
ranges from different environments into a single IMF contributes to an
uncertainty in the observed 
shape of the IMF \citep{massey98,kroupa02,chabrier03,lada03}.
As shown in Figure~5 of \citet{kroupa02}, estimates of the slope in
the range 0.8--$3 \msun$ show a particularly large scatter, because
neither the Galactic field star or young cluster
methods is very effective in this region.  Kroupa notes that the low
and high mass regions of the IMF are observationally almost disjoint,
owing to the lack of overlap between the mass ranges of the most
accurate methods.  Thus if anything, the hatched regions in 
Figures \ref{fig.imfplot}a and \ref{fig.imfplot}c are too restrictive
in the region around $1 \Msun$, making a paunchy IMF more plausible.
For example, \citet{sirianni00}
used HST photometry of the large star cluster 30 Dor in the LMC
to obtain isochrones in the colour-magnitude diagram.
Their resulting IMF has a slope of $1.28 \pm 0.03$ for $M>2.1 \msun$,
and $0.27 \pm 0.08$ for $M < 2.1 \msun$, at least down to $1.3 \msun$ where
the noise becomes large.  This is tentative support for a paunchy IMF, at
least in some active star-forming environments, which may be more
representative of the main contributors to the cosmic SFH.

Interestingly, \citet{kroupa01} finds evidence for a larger number of
low-mass stars in the {\em present-day} IMF than in the {\em mean
Galactic} IMF, which is dominated by stars formed within the last
$5 \Gyr$.  This is tentative evidence that the 
IMF has been increasingly weighted to stars above $1 \msun$ in the past, and
it would not be surprising if the trend continued to the global population
of stars, which is still older on average than the local Galactic disk.

Let us now consider an IMF that varies systematically with cosmic time.
There are both theoretical and
observational motivations for considering a top-heavy IMF for
high-redshift or high-SFR galaxies.  Problems that this has been
invoked to solve include gas consumption timescales in starbursts
\citep{rieke93}, systematic variations of metallicity and $M/L$ ratios
in ellipticals \citep{zepf96}, and large iron abundances in clusters
\citep{zepf96,maoz04}.
It has often been suggested that in extreme star formation
environments, the formation of low-mass stars should be suppressed either by
the global increased temperature of the medium, which basically raises the
Jeans mass, or by feedback from high-mass
stars \citep{massey98,larson98}. However,  proving this by observing low
and high mass stars together in the same environment is very
difficult.  Simulations suggest that primordial metal-free stars, at
least, form with an extremely top-heavy IMF (with a minimum mass
$\tsim 100 \msun$; \citealp{abel02}), while simulations of starburst
galaxies also suggest a top-heavy IMF \citep{padoan97}.  The Milky Way
IMF shows two characteristic mass scales, at $\tsim 0.01$ and $\tsim
0.5 \msun$ \citep{kroupa01}, whose origin is not well understood.  It
would not be surprising if these characteristic masses shifted with
redshift, metallicity, or SFR.

As a toy model to investigate this idea,
we suppose that there are both ``normal'' and
``rapid'' modes of star formation and associate the latter with
starbursting galaxies, making it much more important at high
redshifts (cf.\ \citealp{lagache03}).  We assume both modes have
a Kennicutt IMF, but sharply truncate it in the  ``rapid'' mode below $3
\msun$.  We assume that the normal mode contributes a fixed amount of
star formation at all redshifts, which is 90\% of the star formation rate today,
and that the rapid mode contributes the remainder. In this model the
rapid mode quickly becomes the dominant form as the redshift increases.
At very high redshifts, where the SFR rate falls below  90\% of the rate today,
we just assume that all the star formation is in the normal mode.
Each component is scaled
according to its UV luminosity at $10^8 \yr$ to match the specified
fraction of the best-fit SFH.

We show the time-averaged IMF produced by this toy model in
Figure~\ref{fig.imfplot}c, and its resulting location in the
$\rho_K$-$\jbg$ plane in Figure~\ref{fig.imfplot}d.  Clearly, this
type of model can produce large background levels, especially if the
calibration of the SFH is poorly known at high redshift.  We note that
from an energetic standpoint this is an inefficient solution, since
much of the large energy output at early times is redshifted away.
It would be more efficient to make the top-heavy mode dominant at {\it
  low} redshift, but this would then lead to conflicts with
local observational constraints on the IMF.  This particular model can be
described as paunchy, since it has a sudden jump at $3 \msun$.

We conclude that a paunchy or top-heavy IMF is observationally
allowed, and might indeed be required to solve the background problem,
depending on how future observations sort themselves out.  The consequences of a
top-heavy IMF have been frequently considered.  A paunchy IMF also has
numerous consequences, which are similar in some ways and different in
others to a top-heavy IMF.  For a given SFH,
it increases the number of F, A, and late B main-sequence stars
and the number of bright giants.  This will increase the
optical brightness of galaxies relative to the IR, or raise the
specific SFR that is inferred assuming a standard IMF.
Currently, the SFH deduced from the fossil record is significantly
higher than direct SFH measurements at lookback times of $\tsim 5 \Gyr$,
and significantly lower at $\lesssim 1 \Gyr$ and $\gtrsim 8 \Gyr$
\citep{heavens04}.  This suggests that there are fewer low-mass and
high-mass stars than one expects from simply integrating the 
observed SFH, assuming a standard IMF.  Assuming a paunchy IMF would help
solve this discrepancy.

With a paunchy IMF, post-starburst ``E$+$A'' galaxies will appear more
frequently.  The cosmic luminosity density will also be modified, and it would
be worthwhile to repeat the analysis of \citet{baldry03}, with a break
at intermediate stellar masses, to see whether the high-mass slope we
have assumed for the paunchy form could be constrained.  The passive
evolution of galaxies would appear more rapid with a paunchy IMF, so that
the apparent mass function of high-redshift red-sequence galaxies would shift 
to higher masses than if one assumed a standard IMF.
The intermediate-mass stars will leave
large numbers of white dwarfs, amounting to $\tsim 20\%$ of the total
stellar mass, which could be constrained by microlensing surveys.
Finally, the chemical byproducts of intermediate-mass stars are very
different from those of high-mass stars, tending to make
``$\alpha$-suppressed'' abundances. Hence, the analysis of cosmic abundances may
be able to place limits on the shape of this type of IMF.

\section{CONCLUSIONS}
\label{sec.conclude}

We have investigated the relationship between the total amount of
extragalactic background light, the cosmic star formation history,
and the stellar population today.  We find that there is tension,  
and arguably outright conflict, between the
background light and NIR surveys that probe the local stellar mass
density around $1 \msun$, if we assume a standard Salpeter IMF slope.
We have paid particular attention to uncertainty in the star formation
history; we find that it contributes only a small dispersion to the
predicted ratio of background light and NIR light, even though the 
total amount of star formation is highly uncertain.

This ratio, however, is IMF-dependent.  Universal IMFs with steeper
slopes above $1 \msun$, such as \citet{kennicutt83},
\citet{millerscalo79}, or \citet{kroupa93}, are ruled out by the
current minimum energy requirements.  If the higher absolute
background measurements should be substantiated in the future, it
conversely implies that the average IMF is slightly top-heavy, or
more likely (given limits on the observed star formation history),
middle-heavy or ``paunchy'', i.e. rich in 1.5--$4 \msun$
stars compared to the Salpeter IMF.  
Alternative solutions in which non-stellar sources make
large contributions to the background light appear unlikely.

We note that every quantity that one can measure for the stellar
population of galaxies is IMF-dependent, and many secondary
conclusions drawn from the study of galactic stellar populations, e.g.
galaxy merger rates or rates of gas infall, depend in an indirect
way on the IMF.  The extreme difficulty of deriving the stellar IMF
from first principles, the fundamental difficulty of establishing the
IMF above $1 \msun$ from local observations, and the theoretical
suspicion that the IMF varies with environment all suggest that
galactic astronomers may have to deduce the form of the IMF from 
their own data.  For that, we will need to develop good tests involving
closed systems of gas and stars.  Since the Universe is the ultimate
closed system, it seems likely that studies of cosmic-averaged
quantities will be crucial to establish the form of the IMF.  The
study of cosmic star formation and background light in this paper is a
step in that direction.

\vskip+2mm

We thank Martin Weinberg, Andrew Hopkins, Harry Ferguson, Eric Bell, 
Mauro Giavalisco, Laura Silva, Ben Panter, and Steve Schneider for 
useful conversations.
This research was supported by NASA grants NAG5-13102, NAGS-13308, and
NNG04GK68G, and NSF grant AST-0205969.


\begin{thebibliography}{}

\bibitem[Abel, Bryan, \& Norman(2002)]{abel02} Abel, T.,
Bryan, G.~L., \& Norman, M.~L.\ 2002, Science, 295, 93

\bibitem[Aharonian(2001)]{aharonian01} Aharonian, F.~A.\ 2001, 
International Cosmic Ray Conference, 27, 250 (astro-ph/0112314)

\bibitem[Aharonian et al.(2002)]{aharonian02}
Aharonian, F.~et al.\ 2002, \aap, 384, L23 

\bibitem[Aharonian et al.(2005)]{aharonian05}
Aharonian, F.~et al.\ 2005, Nature, accepted (astro-ph/0508073)

\bibitem[Aller \& Richstone(2002)]{aller02}
Aller, M.~C., \& Richstone, D.\ 2002, \aj, 124, 3035


\bibitem[Baldry \& Glazebrook(2003)]{baldry03} Baldry, I.~K., \& 
Glazebrook, K.\ 2003, \apj, 593, 258

\bibitem[Barger, Cowie, \& Richards(2000)]{barger00} Barger, 
A.~J., Cowie, L.~L., \& Richards, E.~A.\ 2000, \aj, 119, 2092

\bibitem[Bell \& de Jong(2001)]{bell01}
Bell, E.~F., \& de Jong, R.~S.\ 2001, \apj, 550, 212 

\bibitem[Bell et al.(2003)]{bell03} Bell, E.~F., McIntosh, 
D.~H., Katz, N., \& Weinberg, M.~D.\ 2003, \apjs, 149, 289 

\bibitem[Bernstein, Freedman, \& Madore(2002a)]{bernstein02a} 
Bernstein, R.~A., Freedman, W.~L., \& Madore, B.~F.\ 2002, \apj, 571, 56 
\bibitem[Bernstein, Freedman, \& Madore(2002b)]{bernstein02c} 
Bernstein, R.~A., Freedman, W.~L., \& Madore, B.~F.\ 2002, \apj, 571, 107 


\bibitem[Bouwens et al.(2006)]{bouwens06} 
Bouwens, R. J., Illingworth, G. D., Blakeslee, J. P., Franx, M. 
2006, ApJ, accepted (astro-ph/0509641)

\bibitem[Brinchmann \& Ellis(2000)]{brinchmann00} Brinchmann, J., \& 
Ellis, R.~S.\ 2000, \apjl, 536, L77 

\bibitem[Bruzual \& Charlot(2003)]{bruzual03} Bruzual, G., \& 
Charlot, S.\ 2003, \mnras, 344, 1000 

\bibitem[Bullock et al.(2000)]{bullock00} Bullock, J.~S., 
Kravtsov, A.~V., \& Weinberg, D.~H.\ 2000, \apj, 539, 517 

\bibitem[Calzetti(1999)]{calzetti99} Calzetti, D.\ 1999, \apss, 266, 243 
(astro-ph/9902107)

\bibitem[Cambr{\'e}sy et al.(2001)]{cambresy01} Cambr{\'e}sy, L., 
Reach, W.~T., Beichman, C.~A., \& Jarrett, T.~H.\ 2001, \apj, 555, 563 

\bibitem[Chabrier(2003)]{chabrier03} Chabrier, G.\ 2003, \pasp, 
115, 763

\bibitem[Chapman et al.(2005)]{chapman05} Chapman, S.~C., Blain, 
A.~W., Smail, I., \& Ivison, R.~J.\ 2005, \apj, 622, 772 

\bibitem[Cole et al.(2000)]{cole00} Cole, S., Lacey, C.~G., 
Baugh, C.~M., \& Frenk, C.~S.\ 2000, \mnras, 319, 168 

\bibitem[Cole et al.(2001)]{cole01} Cole, S.~et al.\ 2001, 
\mnras, 326, 255 

\bibitem[Connolly et al.(1997)]{connolly97} Connolly, A.~J., 
Szalay, A.~S., Dickinson, M., Subbarao, M.~U., \& Brunner, R.~J.\ 1997,
\apjl, 486, L11

\bibitem[Dekel \& Birnboim(2006)]{dekel06} Dekel, A., \& 
Birnboim, Y.\ 2006, \mnras, 368, 2

\bibitem[Dickinson et al.(2003)]{dickinson03}
Dickinson, M., Papovich, C., Ferguson, H. C., Budavari, T.
2003, ApJ, 587, 25

\bibitem[Dole et al.(2006)]{dole06} Dole, H., et al.\ 2006, 
\aap, accepted (astro-ph/0603208)

\bibitem[Drory et al.(2004)]{drory04} Drory, N., Bender, R., 
Feulner, G., Hopp, U., Maraston, C., Snigula, J., \& Hill, G.~J.\ 2004, 
\apj, 608, 742 

\bibitem[Drory et al.(2005)]{drory05} Drory, N., Salvato, M., 
Gabasch, A., Bender, R., Hopp, U., Feulner, G., \& Pannella, M.\ 2005, 
\apjl, 619, L131 

\bibitem[Dwek et al.(2005)]{dwek05} Dwek, E., Arendt, R.~G., 
\& Krennrich, F.\ 2005, \apj, 635, 784 

\bibitem[Durrell et al.(2002)]{durrell02} Durrell, P.~R., 
Ciardullo, R., Feldmeier, J.~J., Jacoby, G.~H., \& Sigurdsson, S.\ 2002, 
\apj, 570, 119 

\bibitem[Eke et al.(2005)]{eke05} Eke, V.~R., Baugh, C.~M., 
Cole, S., Frenk, C.~S., King, H.~M., \& Peacock, J.~A.\ 2005, \mnras, 362, 
1233 


\bibitem[Fazio et al.(2004)]{fazio04} Fazio, G.~G., et al.\ 
2004, \apjs, 154, 39 

\bibitem[Fioc \& Rocca-Volmerange(1997)]{fioc97} Fioc, M.\ \& 
Rocca-Volmerange, B.\ 1997, \aap, 326, 950 

\bibitem[Finkbeiner et al.(2000)]{finkbeiner00} Finkbeiner, D.~P., 
Davis, M., \& Schlegel, D.~J.\ 2000, \apj, 544, 81 

\bibitem[Fixsen et al.(1998)]{fixsen98} Fixsen, D.~J., Dwek, E., 
Mather, J.~C., Bennett, C.~L., \& Shafer, R.~A.\ 1998, \apj, 508, 123 

\bibitem[Fontana et al.(2003)]{fontana03} Fontana, A., et al.\ 
2003, \apjl, 594, L9 

\bibitem[Fontana et al.(2004)]{fontana04} Fontana, A., et al.\ 
2004, \aap, 424, 23 

\bibitem[Fukugita, Hogan, \& Peebles(1998)]{fukugita98} Fukugita, 
M., Hogan, C.\ J., \& Peebles, P.\ J.\ E.\ 1998, \apj, 503, 518 

\bibitem[Fukugita \& Peebles(2004)]{fugukita04} Fukugita, M., \& 
Peebles, P.~J.~E.\ 2004, \apj, 616, 643 

\bibitem[Gabasch et al.(2004)]{gabasch04} Gabasch, A., et al.\ 
2004, \apjl, 616, L83 

\bibitem[Gallego et al.(1995)]{gallego95} Gallego, J., Zamorano, J., 
Aragon-Salamanca, A., \& Rego, M.\ 1995, \apjl, 455, L1 

\bibitem[Giavalisco et al.(2004)]{giavalisco04} Giavalisco, M., et 
al.\ 2004, \apjl, 600, L103 

\bibitem[Gispert, Lagache, \& Puget(2000)]{gispert00} Gispert,
R., Lagache, G., \& Puget, J.~L.\ 2000, \aap, 360, 1

\bibitem[Gould, Bahcall, \& Flynn(1996)]{gould96} Gould, A.,
Bahcall, J.~N., \& Flynn, C.\ 1996, \apj, 465, 759

\bibitem[Gnedin(2000)]{gnedin00} 
Gnedin, N.~Y.\ 2000, \apj, 542, 535 

\bibitem[Gorjian et al.(2000)]{gorjian00} Gorjian, V., Wright, 
E.~L., \& Chary, R.~R.\ 2000, \apj, 536, 550

\bibitem[Graham \& Driver(2005)]{graham05} Graham, A.~W., \& 
Driver, S.~P.\ 2005, Publications of the Astronomical Society of Australia, 
22, 118 

\bibitem[Guhathakurta et~al.(2005)]{raja05} Guhathakurta, P., Rich, R. M.,
    Reitzel, D.~B., Cooper, M.~C., Gilbert, K., Majewski, S.~R., Ostheimer,
    J.~C., Geha, M. C., Johnston, K. V., \& Patterson, R. J. 2005, AJ,
    submitted (astro-ph/0406145)

\bibitem[Gwyn \& Hartwick(2005)]{gwyn05} Gwyn, S.~D.~J., \& 
Hartwick, F.~D.~A.\ 2005, \aj, 130, 1337

\bibitem[Hauser et al.(1998)]{hauser98} Hauser, M.~G., et al.\ 
1998, \apj, 508, 25 

\bibitem[Hauser \& Dwek(2001)]{hauser01review} Hauser, M.~G., \& Dwek, 
E.\ 2001, \araa, 39, 249 

\bibitem[Heavens et al.(2004)]{heavens04} Heavens, A., Panter, 
B., Jimenez, R., \& Dunlop, J.\ 2004, \nat, 428, 625

\bibitem[Hopkins et al.(2000)]{hopkins00} Hopkins, A.~M., 
Connolly, A.~J., \& Szalay, A.~S.\ 2000, \aj, 120, 2843 

\bibitem[Hopkins \& Beacom(2006)]{hopkins06}
Hopkins, A.~M., \& Beacom, J. F. 2006, submitted to ApJ, astro-ph/0601463

\bibitem[Hopkins(2004)]{hopkins04} Hopkins, A.~M.\ 2004, \apj, 
615, 209 

\bibitem[Hopkins et al.(2006)]{phopkins06} Hopkins, P. 
Laas, \& co. 2006, ApJS, accepted (astro-ph/0506398)

\bibitem[Jones et al.(2006)]{jones06} Jones, D. H., Peterson, B. A.,
Colless, M., Saunders, W. 2006, accepted to MNRAS (astro-ph/0603609)

\bibitem[Kashlinsky et al.(1996a)]{kashlinsky96a} Kashlinsky, A., 
Mather, J.~C., Odenwald, S., \& Hauser, M.~G.\ 1996, \apj, 470, 681 

\bibitem[Kashlinsky et al.(1996b)]{kashlinsky96b} Kashlinsky, A., 
Mather, J.~C., \& Odenwald, S.\ 1996, \apjl, 473, L9 

\bibitem[Kashlinsky \& Odenwald(2000)]{kashlinsky00} Kashlinsky, A., 
\& Odenwald, S.\ 2000, \apj, 528, 74 

\bibitem[Katz et al.(1992)]{katz92} Katz, N., Hernquist, L., 
\& Weinberg, D.~H.\ 1992, \apjl, 399, L109 

\bibitem[Kauffmann et al.(1999)]{kauffmann99} Kauffmann, G., 
Colberg, J.~M., Diaferio, A., \& White, S.~D.~M.\ 1999, \mnras, 303, 188 

\bibitem[Kennicutt(1983)]{kennicutt83} Kennicutt, R.~C.\ 1983, 
\apj, 272, 54 

\bibitem[Kere{\v s} et al.(2005)]{keres05} Kere{\v s}, D., 
Katz, N., Weinberg, D.~H., \& Dav{\'e}, R.\ 2005, \mnras, 363, 2 

\bibitem[Kroupa et al.(1993)]{kroupa93} Kroupa, P., Tout, C.~A., 
\& Gilmore, G.\ 1993, \mnras, 262, 545 

\bibitem[Kroupa(2001)]{kroupa01} 
Kroupa, P.\ 2001, \mnras, 322, 231 

\bibitem[Kroupa(2002)]{kroupa02} Kroupa, P.\ 2002, Science, 295, 82 

\bibitem[Lada \& Lada(2003)]{lada03} Lada, C.~J., \& Lada, 
E.~A.\ 2003, \araa, 41, 57 

\bibitem[Lagache et al.(2000)]{lagache00} Lagache, G., Haffner, 
L.~M., Reynolds, R.~J., \& Tufte, S.~L.\ 2000, \aap, 354, 247 

\bibitem[Lagache et al.(2003)]{lagache03} Lagache, G., Dole, H., 
\& Puget, J.-L.\ 2003, \mnras, 338, 555 

\bibitem[Larson(1998)]{larson98} Larson, R.~B.\ 1998, \mnras,
301, 569

\bibitem[Madau et al.(1996)]{madau96} Madau, P., Ferguson, 
H.~C., Dickinson, M.~E., Giavalisco, M., Steidel, C.~C., \& Fruchter, A.\ 
1996, \mnras, 283, 1388 

\bibitem[Madau \& Pozzetti(2000)]{madau00} Madau, P.\ \& 
Pozzetti, L.\ 2000, \mnras, 312, L9 

\bibitem[Madau et al.(2001)]{madau01} Madau, P., Haardt, F., \& 
Pozzetti, L.\ 2001, IAU Symposium, 204, 359 

\bibitem[Madau \& Silk(2005)]{madau05} Madau, P., \& Silk, J.\ 
2005, \mnras, 359, L37

\bibitem[Marconi et al.(2004)]{marconi04}
Marconi, A., Risaliti, 
G., Gilli, R., Hunt, L.~K., Maiolino, R., \& Salvati, M.\ 2004, \mnras, 
351, 169 

\bibitem[Martin et al.(2005)]{martin05} Martin, D.~C., et al.\ 
2005, \apjl, 619, L59 

\bibitem[Massey(1998)]{massey98} Massey, P.\ 1998, ASP 
Conf.~Ser.~142: The Stellar Initial Mass Function (38th Herstmonceux 
Conference), 142, 17 

\bibitem[Matsumoto et al.(2005)]{matsumoto05} Matsumoto, T., et 
al.\ 2005, \apj, 626, 31 

\bibitem[Maoz \& Gal-Yam(2004)]{maoz04} Maoz, D., \& Gal-Yam, 
A.\ 2004, \mnras, 347, 951 

\bibitem[Miller \& Scalo(1979)]{millerscalo79} Miller, G.~E., \& 
Scalo, J.~M.\ 1979, \apjs, 41, 513 

\bibitem[Mouhcine et al.(2005)]{mouhcine05} Mouhcine, M., 
Ferguson, H.~C., Rich, R.~M., Brown, T.~M., \& Smith, T.~E.\ 2005, \apj, 
633, 821 

\bibitem[Nagamine et al.(2000)]{nagamine00} Nagamine, K., Cen, R., 
\& Ostriker, J.~P.\ 2000, \apj, 541, 25 

\bibitem[Nagamine et al.(2006)]{nagamine06} Nagamine, K., 
Ostriker, J.~P., Cen, R., \& Fukugita, M.\ 2000, ApJ, submitted (astro-ph/0603257)

\bibitem[Padoan, Nordlund, \& Jones(1997)]{padoan97} Padoan, P.,
Nordlund, A., \& Jones, B.~J.~T.\ 1997, \mnras, 288, 145

\bibitem[Okamura et al.(2002)]{okamura02} Okamura, S.~et al.\ 
2002, \pasj, 54, 883 

\bibitem[Panter et al.(2003)]{panter03} Panter, B., Heavens, 
A.~F., \& Jimenez, R.\ 2003, \mnras, 343, 1145 

\bibitem[Panter et al.(2004)]{panter04} Panter, B., Heavens, 
A.~F., \& Jimenez, R.\ 2004, \mnras, 355, 764 

\bibitem[Papovich et al.(2004)]{papovich04} Papovich, C., et al.\ 
2004, \apjs, 154, 70 

\bibitem[P{\'e}rez-Gonz{\'a}lez et al.(2003)]{perez03} 
P{\'e}rez-Gonz{\'a}lez, P.~G., Zamorano, J., Gallego, J., 
Arag{\'o}n-Salamanca, A., \& Gil de Paz, A.\ 2003, \apj, 591, 827 

\bibitem[P{\'e}rez-Gonz{\'a}lez et al.(2005)]{perez05} 
P{\'e}rez-Gonz{\'a}lez, P.~G., et al.\ 2005, \apj, 630, 82 

\bibitem[Pozzetti \& Madau(2001)]{pozzetti01}
Pozzetti, L., \& Madau, P.
2001, IAU Symposium (astro-ph/0011359)

\bibitem[Primack et al.(2005)]{primack05} Primack, J.~R., 
Bullock, J.~S., \& Somerville, R.~S.\ 2005, AIP Conf.~Proc.~745: High 
Energy Gamma-Ray Astronomy, 745, 23 

\bibitem[Salpeter(1955)]{salpeter55} Salpeter, E.~E.\ 1955, \apj, 
121, 161 

\bibitem[Salvaterra \& Ferrara(2006)]{salvaterra06} Salvaterra, R., 
\& Ferrara, A.\ 2006, \mnras, 216 

\bibitem[Schiminovich et al.(2005)]{schiminovich05} Schiminovich, D., 
et al.\ 2005, \apjl, 619, L47 

\bibitem[Rieke, Loken, Rieke, \& Tamblyn(1993)]{rieke93} Rieke,
G.~H., Loken, K., Rieke, M.~J., \& Tamblyn, P.\ 1993, \apj, 412, 99

\bibitem[Santos et al.(2002)]{santos02} Santos, M.~R., Bromm, 
V., \& Kamionkowski, M.\ 2002, \mnras, 336, 1082 

\bibitem[Sawicki \& Thompson(2005)]{sawicki05} Sawicki, M., \& 
Thompson, D.\ 2005, \apj, 635, 100 

\bibitem[Silva et al.(2004)]{silva04} Silva, L., Maiolino, R., 
\& Granato, G.~L.\ 2004, \mnras, 355, 973

\bibitem[Sirianni et al.(2000)]{sirianni00} Sirianni, M., Nota, 
A., Leitherer, C., De Marchi, G., \& Clampin, M.\ 2000, \apj, 533, 203 

\bibitem[Soltan(1982)]{soltan82}
Soltan, A. 1982, \mnras, 200, 115 

\bibitem[Somerville, Primack, \& Faber(2001)]{somerville01}
Somerville, R.~S., Primack, J.~R., \& Faber, S.~M.\ 2001, \mnras, 320, 504

\bibitem[Spergel et al.(2006)]{spergel06} Spergel, D. N., et al.\  
2006, submitted to ApJ (astro-ph/0603449)

\bibitem[Springel \& Hernquist(2003)]{springel03} Springel, V.~\&
Hernquist, L.\ 2003, \mnras, 339, 312

\bibitem[Springel et al.(2005)]{springel05} Springel, V., Di 
Matteo, T., \& Hernquist, L.\ 2005, \apjl, 620, L79 

\bibitem[Steidel et al.(1999)]{steidel99} Steidel, C.~C., 
Adelberger, K.~L., Giavalisco, M., Dickinson, M., \& Pettini, M.\ 1999,
\apj, 519, 1

\bibitem[Strauss et al.(2002)]{strauss02}
Strauss, M.A., et al.\ 2002, \aj, 124, 1810

\bibitem[Sullivan et al.(2000)]{sullivan00} Sullivan, M., Treyer, 
M.\ A., Ellis, R.\ S., Bridges, T.\ J., Milliard, B., \& Donas, J.\ ; 2000, 
\mnras, 312, 442 

\bibitem[Tapken al.(2006)]{tapken06} Tapken, C., et al.\ 
2006, \aap, accepted (astro-ph/0604262)

\bibitem[Treister et al.(2006)]{treister06} Treister, E., et al.\ 
2006, \apj, 640, 603 

\bibitem[Tresse \& Maddox(1998)]{tresse98} Tresse, L.\ \& 
Maddox, S.\ J.\ 1998, \apj, 495, 691 

\bibitem[Tresse et al.(2002)]{tresse02} Tresse, L., Maddox, 
S.~J., Le F{\`e}vre, O., \& Cuby, J.-G.\ 2002, \mnras, 337, 369 

\bibitem[Tumlinson et al.(2004)]{tumlinson04} Tumlinson, J., 
Venkatesan, A., \& Shull, J.~M.\ 2004, \apj, 612, 602 

\bibitem[Webb et al.(2006)]{webb06} Webb, T.~M.~A., et al.\ 
2006, \apjl, 636, L17 

\bibitem[Wolf et al.(2003)]{wolf03} Wolf, C., Meisenheimer, 
K., Rix, H.-W., Borch, A., Dye, S., \& Kleinheinrich, M.\ 2003, \aap, 401, 73 

\bibitem[Wright(2001)]{wright01} Wright, E.~L.\ 2001, \apj, 553, 538

\bibitem[Wright(2004)]{wright04} Wright, E.~L.\ 2004, New 
Astronomy Review, 48, 465 

\bibitem[Wolfe et al.(2003)]{wolfe03} Wolfe, A.~M., Gawiser, 
E., \& Prochaska, J.~X.\ 2003, \apj, 593, 235 

\bibitem[Worthey(1994)]{worthey94} Worthey, G.\ 1994, \apjs, 95, 107 

\bibitem[Wright \& Reese(2000)]{wright00} 
Wright, E.~L., \& Reese, E.~D.\ 2000, \apj, 545, 43 

\bibitem[Yan et al.(1999)]{yan99} Yan, L., McCarthy, P.~J., 
Freudling, W., Teplitz, H.~I., Malumuth, E.~M., Weymann, R.~J., \& Malkan,
M.~A.\ 1999, \apjl, 519, L47

\bibitem[Yun, Reddy, \& Condon(2001)]{yun01} Yun, M.~S.,
Reddy, N.~A., \& Condon, J.~J.\ 2001, \apj, 554, 803

\bibitem[Zepf \& Silk(1996)]{zepf96} Zepf, S.~E.~\& Silk, J.\
1996, \apj, 466, 114

\end{thebibliography}
\end{document}